\documentclass[aps,prf,twocolumn,groupedaddress,amsmath,amssymb,longbibliography]{revtex4-1}
\usepackage{graphicx}  
\usepackage{dcolumn}   
\usepackage{bm}        
\usepackage{verbatim}   

\usepackage{braket}
\usepackage{pstricks}
\usepackage{epsfig}

\newcommand*{\Ta}{{\rm Ta}}

\newcommand*{\Ray}{{\rm Ra}}
\newcommand*{\Rayc}{{\rm Ra_c}}
\newcommand{\Pra}{{\rm Pr}}

\newcommand{\er}{{\bf\hat e}_r}
\newcommand{\ve}{{\mathbf{v}}}

\begin{document}

\title{The onset of low Prandtl number thermal convection in thin
  spherical shells}

\author{F. Garcia}
\affiliation{
Anton Pannekoek Institute for Astronomy, University of Amsterdam, Postbus 94249, 1090 GE Amsterdam, The Netherlands  
}
\author{F.~R.~N. Chambers}
\affiliation{
Anton Pannekoek Institute for Astronomy, University of Amsterdam, Postbus 94249, 1090 GE Amsterdam, The Netherlands    
}
\author{A.~L. Watts}
\affiliation{
Anton Pannekoek Institute for Astronomy, University of Amsterdam, Postbus 94249, 1090 GE Amsterdam, The Netherlands    
}

\date{\today}
\begin{abstract}


This study considers the onset of stress-free Boussinesq thermal
convection in rotating spherical shells with aspect ratio
$\eta=r_i/r_o=0.9$ ($r_i$ and $r_o$ being the inner and outer radius),
Prandtl numbers $\Pr\in[10^{-4},10^{-1}]$, and Taylor numbers
$\Ta\in[10^{4},10^{12}]$.  We are particularly interested in the form
of the convective cell pattern that develops, and in its time scales,
since this may have observational consequences. For a fixed
$\Ta<10^{9}$ and by decreasing $\Pr$ from 0.1 to $10^{-4}$ a
transition between spiralling columnar (SC) and equatorially-attached
(EA) modes, and a transition between EA and equatorially antisymmetric
or symmetric polar (AP/SP) weakly multicellular modes are found. The
latter modes are preferred at very low $\Pr$. Surprisingly, for
$\Ta>3\times 10^{9}$ the unicellular polar modes become also preferred
at moderate $\Pr\sim10^{-2}$ because two new transition curves between
EA and AP/SP and between AP/SP and SC modes are born at a triple-point
bifurcation. The dependence on $\Pr$ and $\Ta$ of the transitions is
studied to estimate the type of modes, and their critical parameters,
preferred at different stellar regimes.

 \end{abstract}

\maketitle

\section{Introduction}

Convection is believed to occur in many geophysical and astrophysical
objects such as planets and stars.  Compressible convection develops,
for example in main sequence stars (including our Sun), and during
thermonuclear flashes in Asymptotic Giant Branch stars and in the
accreted oceans of white dwarfs and neutron stars. These convective
regions may be formed by very thin spherical layers ($r_i>0.8r_o$) of
Helium or Hydrogen which are subject to the influence of strong
temperature gradients and rotation. From nuclear physics theory the
physical properties, such as kinematic viscosity or thermal
conductivity, can be estimated and may give rise to very low Prandtl
and large Taylor numbers (dimensionless numbers characterising the
relative importance of viscous (momentum) diffusivity to thermal
diffusivity, and rotational and viscous forces, respectively). This
parameter regime, in combination with very thin spherical shells,
makes the study of convection extremely challenging, even in the
incompressible case (Boussinesq).

The study of thermal convection is important because it represents a
common mechanism for transporting energy. In the case of rotating
planets and stars it is also essential to maintain their magnetic
fields via the dynamo effect~\cite{DoSo07} or to explain the
differential rotation observed in the sun~\cite{Bus70b} or in the
major planets~\cite{Chr02}. The basic ingredients occuring in these
situations are convection, rotation and spherical geometry. Due to its
relevance, thermal convection in rotating spherical geometries has
been widely studied using numerical, analytic and experimental
approaches. The reviews~\cite{Jon07} and~\cite{Gil00} give a nice
description of them, with application to the Earth's outer core, or
convective stellar interiors, respectively.

One of the most basic steps in the field is the study of the onset of
convection in rotating spherical shells at large Taylor numbers,
$\Ta$. This is important because it reveals the convective patterns
and the critical parameters when the conductive or basic state becomes
unstable in a regime of astrophysical interest. Much work has been
done since~\cite{Rob68,Bus70a} stated the nonaxisymmetric nature of
the instability giving rise to waves travelling in the azimuthal
direction. The previous theories were successively
improved~\cite{Sow77,Yan92,JSM00}, and finally~\cite{DSJJC04}
completed the asymptotic theory for the onset of spiralling columnar
convection in spherical shells that are differentially heated.

Numerical studies~\cite{ZhBu87,Zha92} have shown the relevance of the
Prandtl number, $\Pr$, to the onset of convection. Spiralling columnar
(SC) patterns are preferred at moderate and large $\Pr$ while
convection is trapped in an equatorial band near the outer boundary
when $\Pr$ is sufficiently small (equatorially attached modes,
EA). The study of low Prandtl number fluids is important because they
are likely to occur in stellar convective layers~\cite{Mas91}, where
thermal diffusion dominates and kinematic viscosities are very
low. This is illustrated, for example, in~\cite{GMBMM15}, with the
help of nuclear physics theories, for stars at various evolutionary
stages. For instance in the case of main-sequence stars, where the
stellar material is nondegenerate, thermal diffusivity was
approximated by its radiative contribution, and for the viscosity the
collisional contribution was also included. With the estimates of
$\nu$ and $\kappa$ they obtanied $\Pr<10^{-3}$ for a sequence of ten
different stellar models.

EA modes were described in~\cite{Zha93,Zha94} as a solutions of the
Poincar\'e equation with stress-free boundary condition and low $\Pr$
in rapidly rotating spheres. They are thermo-inertial waves with
azimuthal symmetry $m>0$ travelling in the azimuthal direction,
eastward but also westward for sufficiently low $\Pr$, and trapped in
a thin equatorial band of characteristic latitudinal radius
$(2/m)^{1/2}$. Several theoretical studies~\cite{ZELB01,BuSi04} have
appeared in the last years focused on these waves, which are
characteristic of low $\Pr$ numbers. By expanding the EA modes as a
single and the SC modes as a superposition of quasi-geostrophic modes,
the inertial and convective instability problems were
unified~\cite{ZhLi04,ZLB07b} for $0\le \Pr \Ta^{1/2}<\infty$ either
with the stress-free or nonslip condition. Very recently, in the
zero-Prandtl limit, a torsional axisymmetric and equatorially
antisymmetric mode have been found numerically~\cite{SGN16} and the
asymptotic theory has been developed~\cite{ZLK17}.

Although most of the asymptotic studies agree that the preferred mode
of convective instabilities in rapidly rotating spheres is
equatorially symmetric, there exist some regions in $(\Ta,\Pr)$ parameter
space where equatorially antisymmetric modes may be
preferred~\cite{Zha93,NGS08,GSN08}. The latter study showed that at
high $\Ta$ antisymmetric convection is confined inside a cylinder
tangent to the inner sphere at the equator (antisymmetric polar mode,
AP) in contrast with the equatorially antisymmetric modes of EA type
found in~\cite{Zha95}. In addition, at the same range of
parameters~\cite{GSN08} found that the preferred mode can also be of
polar type, but equatorially symmetric (symmetric polar mode, SP).

The case of very thin shells ($\eta>0.8$) is especially relevant for
convection occurring in the interiors or accreted ocean layers of
stars, and strongly different from the case of thick shells
($\eta<0.3$). In first place, it is numerically challenging because
the azimuthal length scale is strongly decreased. Most of the studies
mentioned above consider a full sphere or thick spherical shell, but
only a few of them address thin shells: ~\cite{Zha92,DSJJC04} for
instance, address the problem in the case of relatively thin shells
with $r_i=0.65r_o$, and~\cite{ABW97} consider a thinner shell with
$r_i=0.8r_o$ but by assuming equatorial symmetry of the flow. In the
study~\cite{AHA04}, the effect of increasing the inner radius was
studied in depth in the case of SC convection up to $r_i=0.92r_o$
providing an estimation of the dependence of the critical Rayleigh
number on $r_i/r_o$ showing that the critical azimuthal wave number
increases proportionally. The small azimuthal length scale of the
eigenfunctions imposed severe numerical restrictions, and the previous
study only considered $\Pr=1$ and $\Ta$ up to $10^8$. A second
difference of the very thin shell case is that the number of relevant
marginal stability curves becomes large, and they are closer than for
smaller aspect ratios.  Because of the numerical constraints, a
simplified model of a rotating cylindrical annulus in the small-gap
limit has been adopted in the past~\cite{BuOr86}, but without
including spherical curvature spiralling convection is
impeded~\cite{Zha92}. In addition at $\Pr=7$, multiarmed spiral waves
were found in~\cite{LLCZ10} only for the thin shell case, in the
slowly rotating regime.

This paper is devoted to the numerical computation of the onset of
stress-free convection at small $\Pr$ in a thin spherical rotating
layer. The dependence of the critical parameters on $\Ta$ and $\Pr$ is
addressed by an exhaustive exploration of the parameter space. The
types of instabilities characteristic of different regions of the
parameter space are described, and their transitions traced. We have
found SC, EA and SP or AP modes, the latter being dominant at large
$\Ta$ in two separated regions, at low $\Pr$ (where they are
multicellular) but also unexpectedly at moderate $\Pr$. A triple point
in $(\Ta,\Pr)$ space, where several EA, SC and P modes become
marginally stable, is also found for the first time in a regime of
astrophysical relevance. The possible application of the results to
convection occurring in several types of stars is also discussed.

In \S\ \ref{sec:math} we introduce the formulation of the problem, and
the numerical method used for the linear stability analysis. In
\S\ \ref{sec:Tay} the dependence on $\Ta$ is studied and that on $\Pr$
in \S\ \ref{sec:Pr}. The regions of dominance of the different
instabilities are obtained in \S\ \ref{sec:tran} and their transition
boundaries are analysed. The applicability of the results to stellar
convection is addressed in \S\ \ref{sec:star} and finally, the paper
ends with a brief summary of the results obtained in
\S\ \ref{sec:conc}.

\section{Mathematical model and numerical set up}
\label{sec:math}

Thermal convection of a spherical fluid layer differentially heated,
rotating about an axis of symmetry with constant angular velocity
${\bf \Omega}=\Omega {\mathbf{k}}$, and subject to radial gravity
${\bf g}=-\gamma \mathbf{r}$ (with $\gamma$ being a constant and
$\mathbf{r}=r\er$ the position vector), is considered (at this stage
we assume Newtonian gravity and neglect oblateness due to rotation;
for some of the astrophysical applications that we discuss later,
these assumptions may eventually need to be relaxed).  The mass,
momentum and energy equations are written in the rotating frame of
reference. We use scaled variables, with units of $d=r_{o}-r_{i}$ for
distance, $\nu^2/\gamma\alpha d^4$ for temperature, and $d^2/\nu$ for
time. In the previous definitions $\nu$ is the kinematic viscosity and
$\alpha$ the thermal expansion coefficient.

We assume an incompressible fluid by using the Boussinesq
approximation.  The Boussinesq approximation is a useful
simplification that renders the problem more tractable, and allows us
to relate our results to previous studies that were carried out in the
regime of smaller $\Ta$ and larger $\Pr$.  It will not be entirely
appropriate for all of the astrophysical problems of interest, but is
nonetheless useful as a first step towards the full general problem.
We discuss this in more detail in Section~\ref{sec:star}.
  
The velocity field $\mathbf{v}$ is expressed in terms of toroidal,
$\Psi$, and poloidal, $\Phi$, potentials
\begin{equation}
\mathbf{v}=\bm\nabla\times\left(\Psi
\mathbf{r}\right)+\bm\nabla\times\bm\nabla\times\left(\Phi\mathbf{r}\right).
\end{equation}

The linearised equations for both potentials, and the temperature
perturbation, $\Theta=T-T_c$, from the conduction state (which has
${\bf v}={\bf 0}$ and temperature $T_c\equiv T_c(r)=T_0+\Ray\,
\eta/\Pr(1-\eta)^2r$, with $T_0=T_i-r_o \Delta T/d$ being a reference
temperature and $\Delta T=T_i-T_o>0$ the imposed difference in
temperature between the inner and outer boundaries, with
$r=|\mathbf{r}|$, see ~\cite{Cha81}), are
\begin{subequations}
\begin{align}
\left[(\partial_t -\nabla^2)L_{2} - 2\Ta^{1/2}\, \partial_\varphi \right]\Psi 
  ~=&~-2\Ta^{1/2} \mathcal{Q}\Phi,\label{eq:psi} \\ 
\left[(\partial_t -\nabla^2)L_{2} - 2\Ta^{1/2}\, \partial_\varphi \right]\nabla^2 \Phi 
  +L_{2}\Theta  ~=&~2\Ta^{1/2} \mathcal{Q}\Psi, \label{eq:phi} \\ 
\left(\Pr\partial_t-\nabla^2\right)\Theta-
\Ray\,\eta\,(1-\eta)^{-2}r^{-3} L_{2}\Phi
~=&0. \label{eq:theta}
\end{align}
\end{subequations}

The parameters of the problem are the Rayleigh number $\Ray$, the
Prandtl number $\Pr$, the Taylor number $\Ta$, and the aspect ratio
$\eta$. They are defined by
\begin{equation}
  \Ray=\frac{\gamma\alpha\Delta T d^4}{\kappa\nu},\quad
  \Ta =\frac{\Omega^2 d^4}{\nu^2},\quad
  \Pr=\frac{\nu}{\kappa}, \quad \eta=\frac{r_{i}}{r_{o}},
\label{eq:param}
\end{equation}
where $\kappa$ is the thermal diffusivity. Notice that this
effectively constitutes the Rayleigh-B\'enard problem in a spherical
rotating geometry.

The operators $L_{2}$ and $\mathcal{Q}$ are defined by $L_2\equiv -r^2
\nabla^2+\partial_r(r^2\partial_r)$ and $\mathcal{Q}\equiv
r\cos\theta\nabla^2-(L_2+r\partial_r)(\cos\theta\partial_r-r^{-1}\sin\theta\partial_\theta)$,
$(r,\theta,\varphi)$ being the spherical coordinates, with $\theta$
measuring the colatitude, and $\varphi$ the longitude. When
stress-free perfect thermally conducting boundaries are used
\begin{equation}
\Phi=\partial^2_{rr}\Phi=\partial_r(\Psi/r)=\Theta=0 \quad \text{at} \quad r=r_i, r_o.
\end{equation}

The equations are discretised and integrated as described
in~\cite{NGS08}.  The potentials and the temperature perturbation are
expanded in spherical harmonics in the angular coordinates, and in the
radial direction a collocation method on a Gauss--Lobatto mesh is
used.  The leading eigenvalues are found by means of an algorithm,
based on subspace or Arnoldi iteration (see~\cite{LSY98}), in which
the time-stepping of the linearised equations, which decouple for each
azimuthal wave number $m$, is required. For the time integration high
order implicit-explicit backward differentiation formulas
(IMEX--BDF)~\cite{GNGS10} are used. In the multi-step IMEX
method we treat the discretised version of the operator $\mathcal{Q}$
explicitly in order to minimise storage requirements when solving
linear systems at each time step. To obtain the initial conditions a
fully implicit variable size and variable order (VSVO) high order BDF
method~\cite{GNGS10} is used.

With respect to previous versions, the code use is parallelised in the
spectral space by using OpenMP directives in a similar way as
in~\cite{GDSN14} for the fully nonlinear solver. In the case of the
linear problem, the spherical harmonics mesh is no longer triangular
and thus the parallelism of the code is better. The code is also based
on optimised libraries of matrix-matrix products (dgemm
GOTO~\cite{GoGe08}). The code has been validated~\cite{NGS08} by
reproducing several results previously published, for instance
in~\cite{AHA04}. For the larger Taylor number considered in this
study, the computations use $n_r=30-60$ radial collocation points and
a spherical harmonic truncation parameter of $L=160-600$, depending on
the azimuthal wave number $m$ considered. With these choices we obtain
errors for the critical parameters below $2\%$ when increasing the
resolution up to $n_r=100$ and $L=1000$.

Given a set of $\Ta$, $\Pr$ and $\eta$ and for sufficiently small
$\Ray$ the conductive state is stable against temperature and velocity
field perturbations. At the critical Rayleigh number $\Rayc$, the
perturbations can no longer be dissipated and thus convection sets in,
usually breaking the axisymmetry of the conduction state by imposing
an $m_c$-azimuthal symmetry ($m_c$ is the critical wave number). Then,
according to~\cite{EZK92}, it is a Hopf bifurcation giving rise to a
wave travelling (propagating) in the azimuthal direction.  In the
rotating frame of reference, negative critical frequencies $\omega_c$
give positive drifting (phase) velocities $c=-\omega_c/m_c$ and the
waves drift in the prograde direction (eastward). These waves are also
usually called rotating waves in the context of bifurcation
theory~\cite{GLM00,SGN13} and have temporal dependence
$u(t,r,\theta,\varphi)=\tilde u(r,\theta,\varphi-c t)$ (where
$u=(\Psi_l^m(r_i),\Phi_l^m(r_i),\Theta_l^m(r_i))$ is the vector
containing the values of the spherical harmonic coefficients at the
inner radial collocation points). Then, these solutions are periodic
in time but their azimuthally averaged properties (such as zonal flow)
are constant.  Notice that in an inertial reference frame the waves
have frequencies $\omega_{\text{I}}=\omega_c\pm m_c\Omega$. Typically,
the critical waves maintain the $\mathcal Z_2$ symmetry with respect
to the equatorial plane, i. e., they are symmetric with respect to the
equator, but as it was shown in~\cite{GSN08}, equatorially
antisymmetric solutions can also be preferred depending on the
parameters and thus fixed symmetry codes should be avoided when
exploring the parameter space. As it will be shown in next sections,
antisymmetric critical solutions are typical when moderate and low
$\Pr$ are considered in thin spherical shells.

\section{Taylor number dependence}
\label{sec:Tay}

\begin{figure}[t!]
\begin{center}
\includegraphics[width=0.9\linewidth]{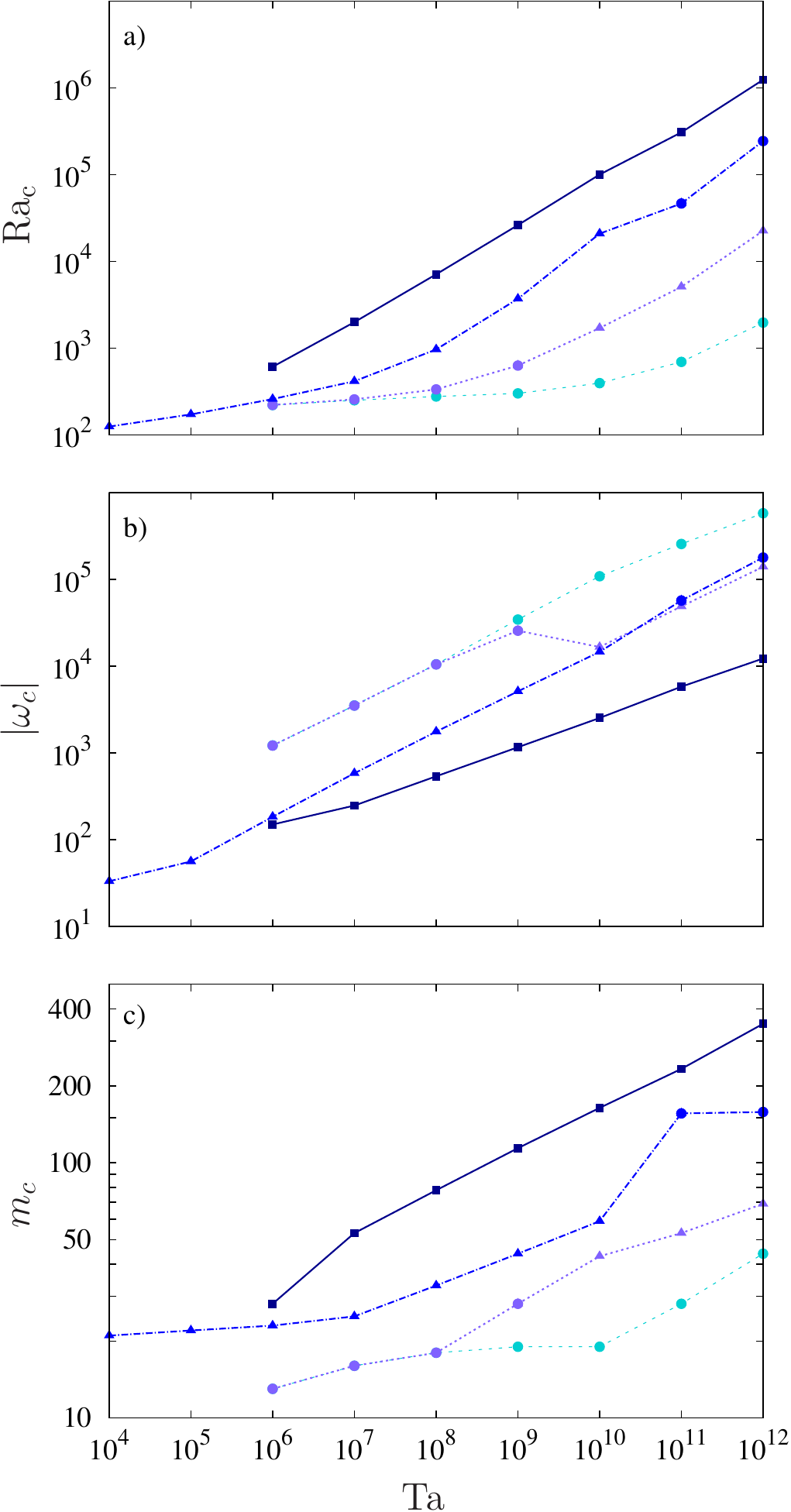}
\end{center}
\caption{(a) Critical Rayleigh numbers $\Rayc$, (b) critical
  drifting frequencies $|\omega_c|$ and (c) critical azimuthal wave
  number $m_c$ versus the Taylor number $\Ta$ for
  $\Pr=10^{-4},10^{-3},10^{-2},10^{-1}$ with dashed, dotted,
  dashed-dotted and solid line, respectively (Blue darkness increasing
  with $\Pr$ online). The symbols mean: $\bullet$ AP/SP mode,
  $\blacktriangle$ EA mode and $\blacksquare$ SC mode. This figure
  roughly explores the $(\Ta,\Pr)$ space as only few points are
  considered. Details for the interchange between EA and AP/SP modes
  (jumps in the figure) at $\Pr=10^{-3},10^{-2}$ are provided in
  Fig.~\ref{fig2}.}
\label{fig1} 
\end{figure}

\begin{figure}[t!]
\begin{center}
\includegraphics[width=0.9\linewidth]{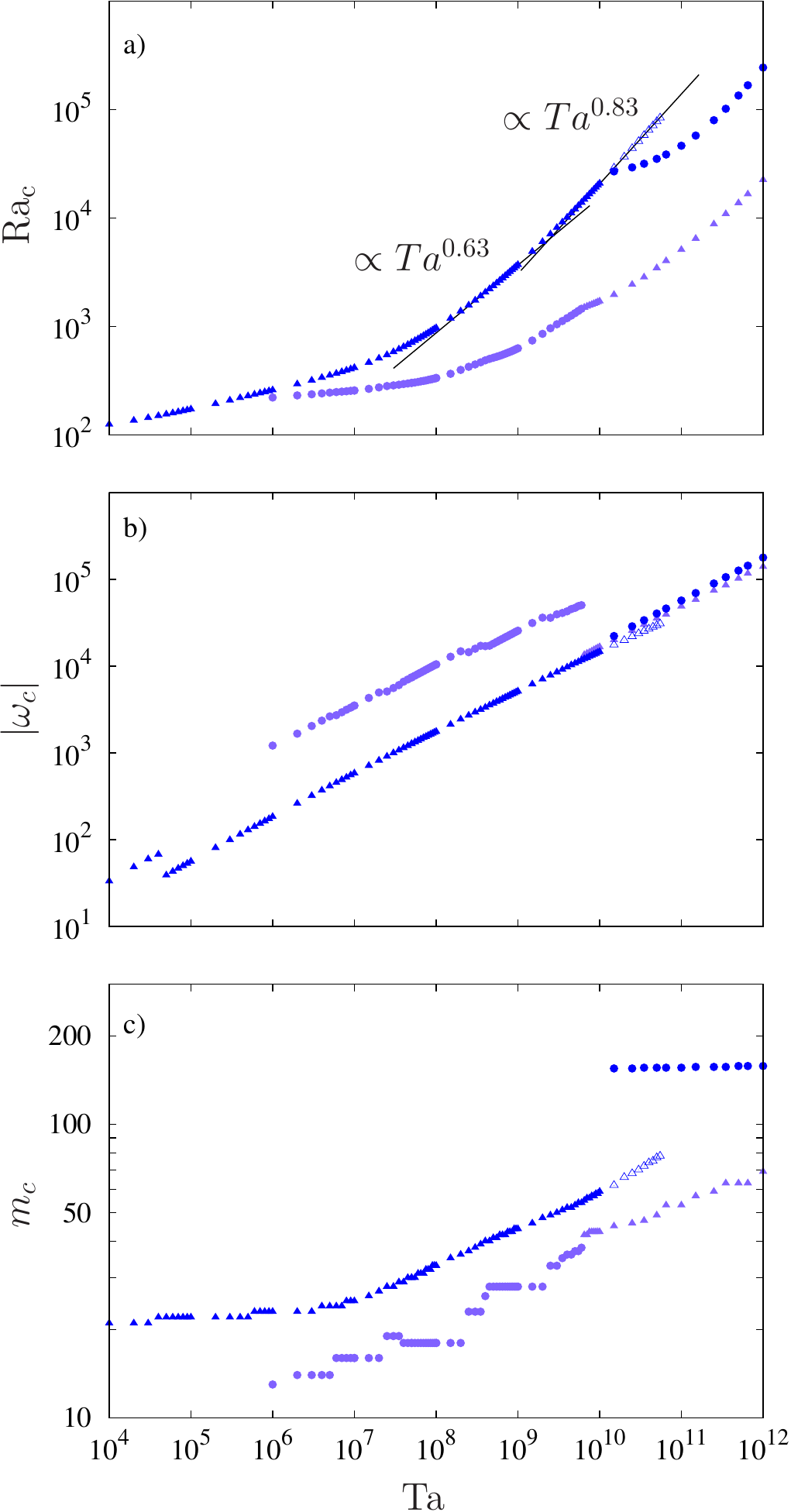}
\end{center}
\caption{(a) Critical Rayleigh numbers $\Rayc$, (b) critical
  drifting frequencies $|\omega_c|$ and (c) critical azimuthal wave
  number $m_c$ versus the Taylor number $\Ta$ for
  $\Pr=10^{-3},10^{-2}$ with light and dark grey (blue online),
  respectively. The symbols mean: $\bullet$ AP/SP mode and
  $\blacktriangle$ EA mode. Full/open points are used for
  dominant/nondominant modes. This figure shows in detail the $\Ta$
  dependence and the interchange between EA and AP/SP modes found only
  at $\Pr=10^{-3},10^{-2}$ and roughly displayed in
  Fig.~\ref{fig1}. In the case of $\Pr=10^{-2}$, nondominant EA modes
  ($\triangle$) are also displayed. The power law fits from the
  results are represented by a solid line.}
\label{fig2} 
\end{figure}

Different types of instabilities have been found by varying the Taylor
number in the range $\Ta\in[10^{4},10^{12}]$ for low and moderate
Prandtl numbers $\Pr=10^{-4},10^{-3},10^{-2},10^{-1}$. The critical
Rayleigh number, $\Ray_c$, drifting frequencies, $|\omega_c|$, and
azimuthal wave number, $m_c$, are displayed versus $\Ta$ in
Fig.~\ref{fig1} and Fig.~\ref{fig2}.  For the largest Prandtl number
studied, $\Pr=0.1$, only spiralling columnar modes, with
$\Rayc=0.2\Ta^{0.57}$, become dominant (see Fig.~\ref{fig1}(a)). This
power law, coming from a fit to the points of the figure, is in close
agreement with the results of~\cite{AHA04}, obtained with $\Pr=1$ and
non-slip boundary conditions also in thin spherical shells, and not so
far from the leading order $2/3$ given by the asymptotic
theories~\cite{Rob65,Bus70a,DSJJC04}. In the case of $|\omega_c|$
(Fig.~\ref{fig1}(b)) or $m_c$ (Fig.~\ref{fig1}(c)) the agreement with
the previous theories~\cite{Rob65,Bus70a,DSJJC04} is quite good giving
$|\omega_c|=\Ta^{0.34}$ and $m_c =4 \Ta^{0.16}$. Spiralling modes
always drift in the prograde direction.

\begin{figure*}
  \begin{center}
    \vspace{0.cm}    
\Large{Antisymmetric Polar (AP) mode\vspace{0.2cm}}
    \begin{tabular}{cccccc}
      \resizebox{!}{31mm}{\includegraphics{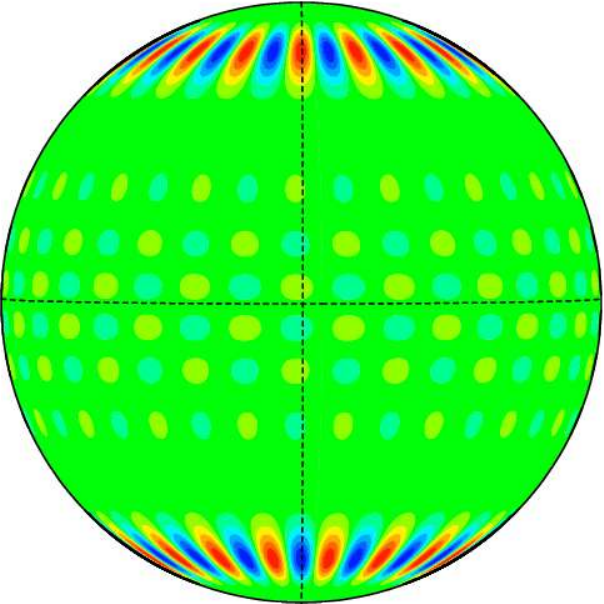}}&
      \resizebox{!}{31mm}{\includegraphics{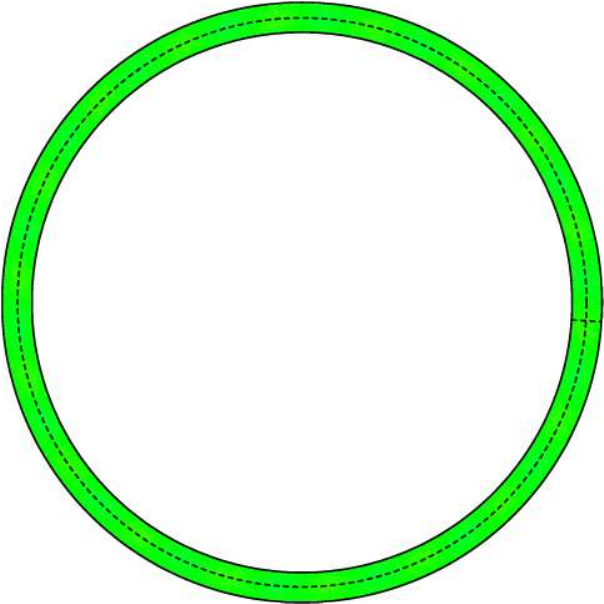}}&
      \resizebox{!}{31mm}{\includegraphics{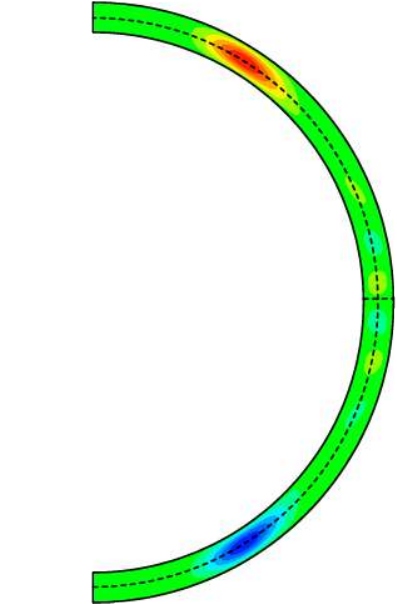}}&
      \resizebox{!}{31mm}{\includegraphics{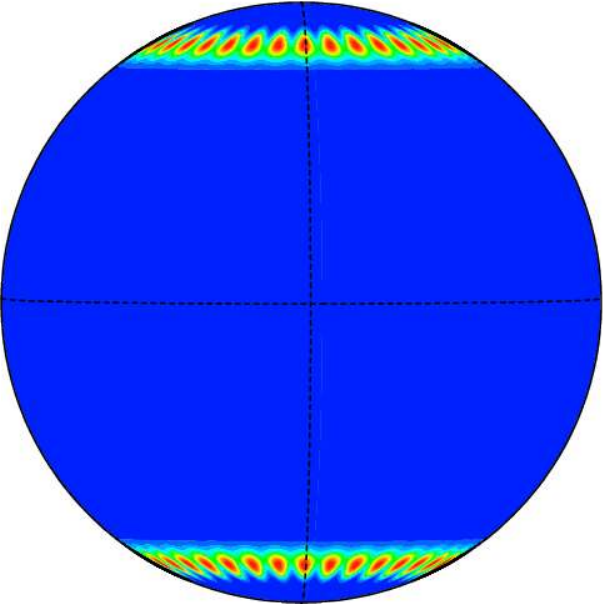}}&
      \resizebox{!}{31mm}{\includegraphics{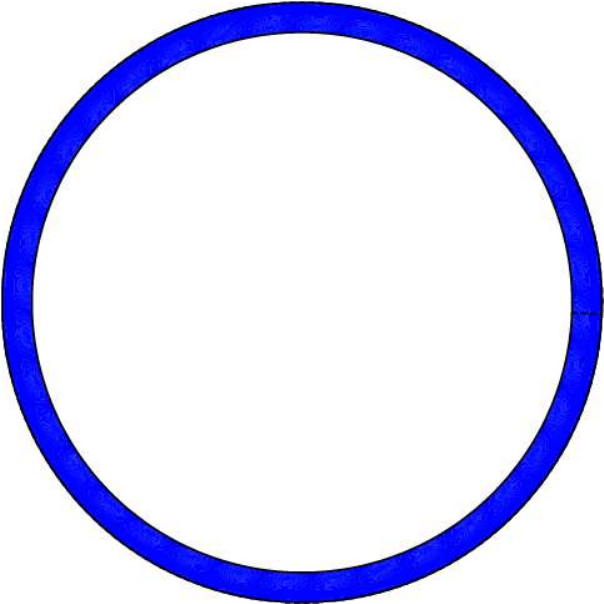}}&
      \resizebox{!}{31mm}{\includegraphics{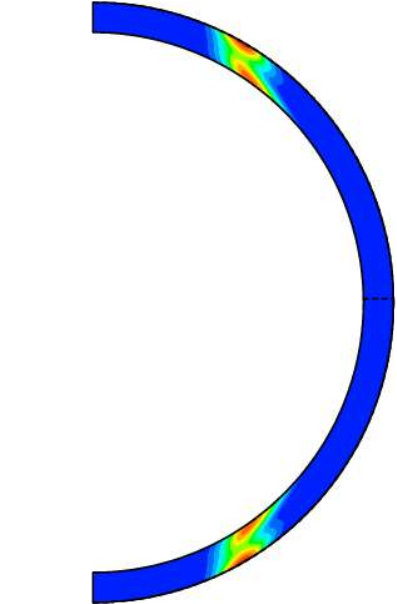}}\\
    \end{tabular}\\
    \vspace{0.4cm}
    \Large{Symmetric Polar (SP) mode\vspace{0.2cm}}
      \begin{tabular}{cccccc}
      \resizebox{!}{31mm}{\includegraphics{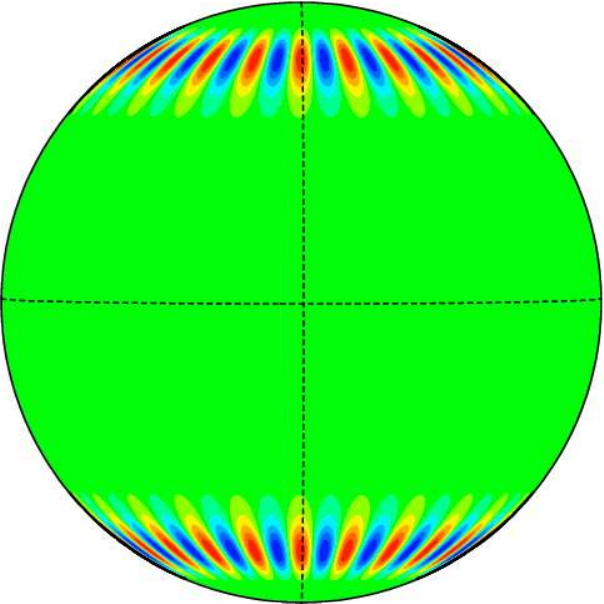}}&
      \resizebox{!}{31mm}{\includegraphics{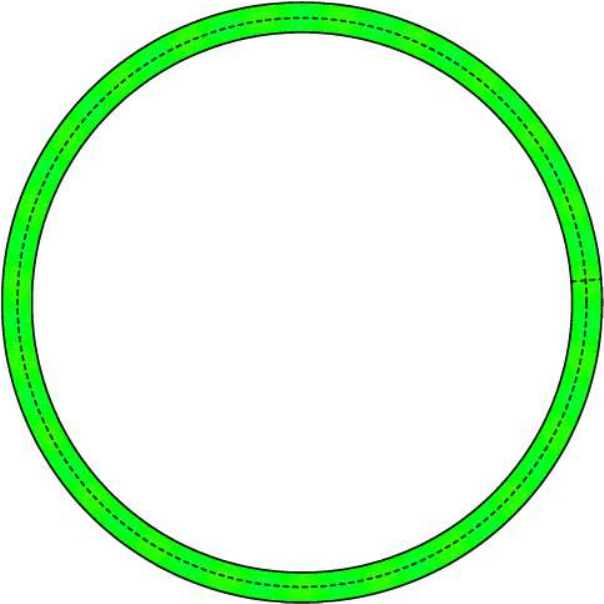}}&
      \resizebox{!}{31mm}{\includegraphics{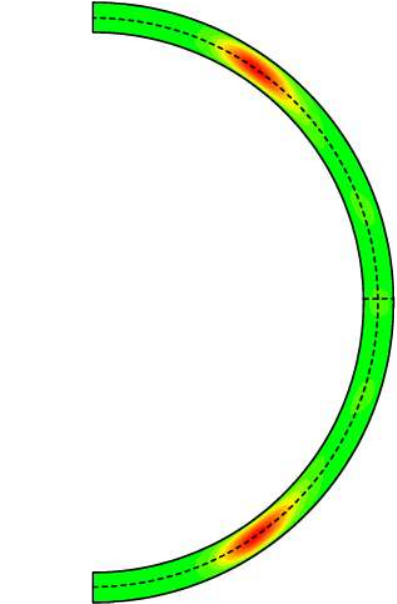}}&
      \resizebox{!}{31mm}{\includegraphics{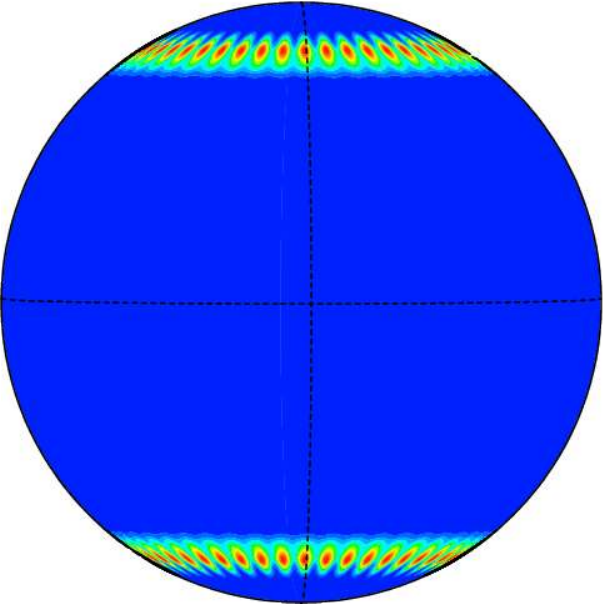}}&
      \resizebox{!}{31mm}{\includegraphics{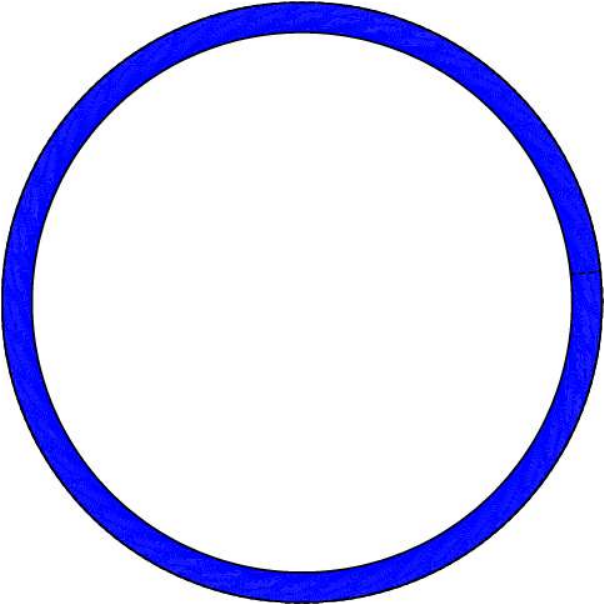}}&
      \resizebox{!}{31mm}{\includegraphics{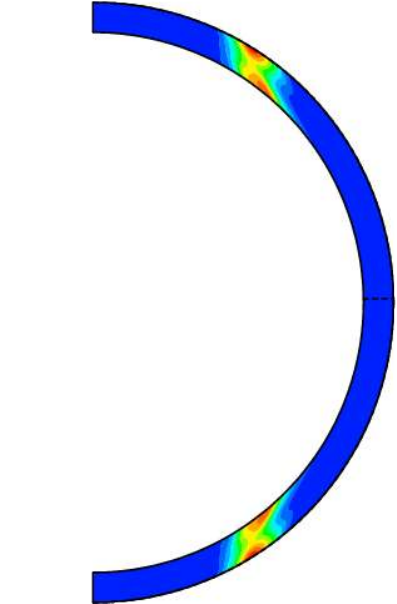}}\\
      \end{tabular}\\
     \vspace{0.4cm}
    \Large{Equatorially Attached (EA) mode\vspace{0.2cm}}
    \begin{tabular}{cccccc}
      \resizebox{!}{31mm}{\includegraphics{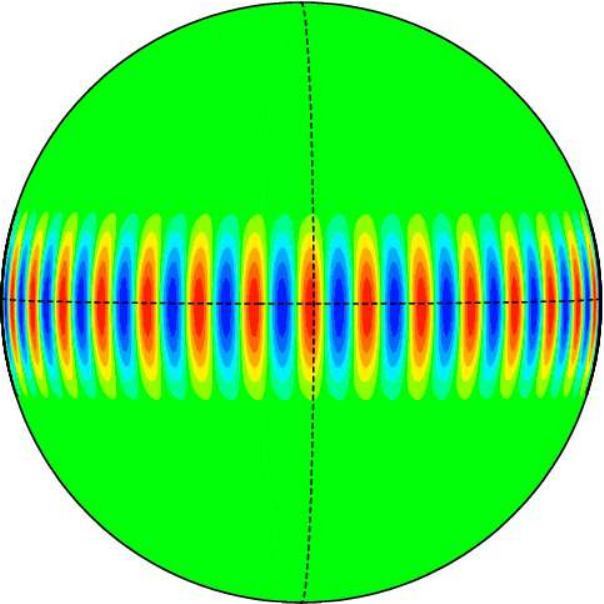}}&
      \resizebox{!}{31mm}{\includegraphics{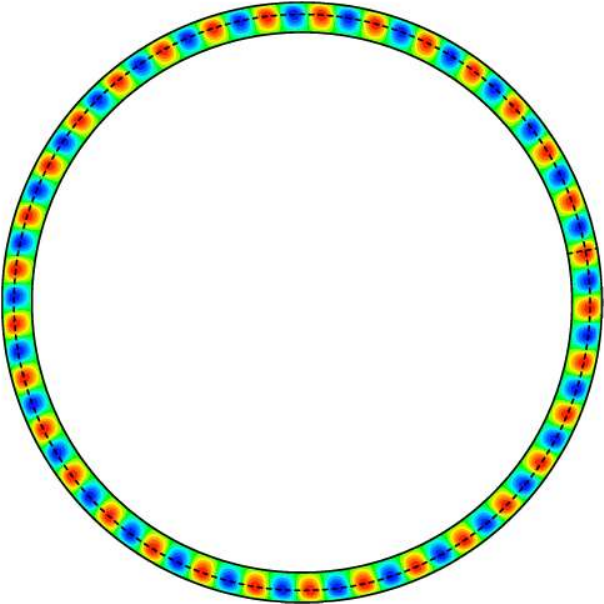}}&
      \resizebox{!}{31mm}{\includegraphics{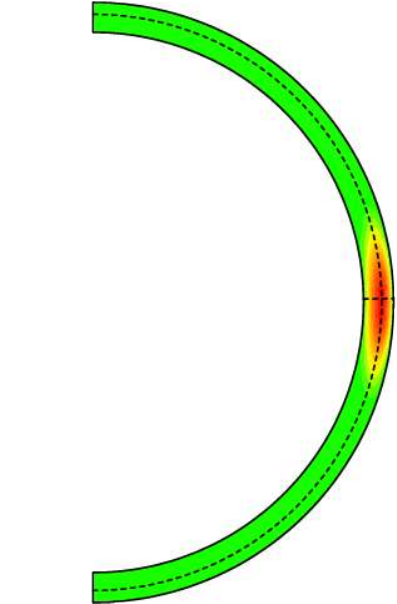}}&
      \resizebox{!}{31mm}{\includegraphics{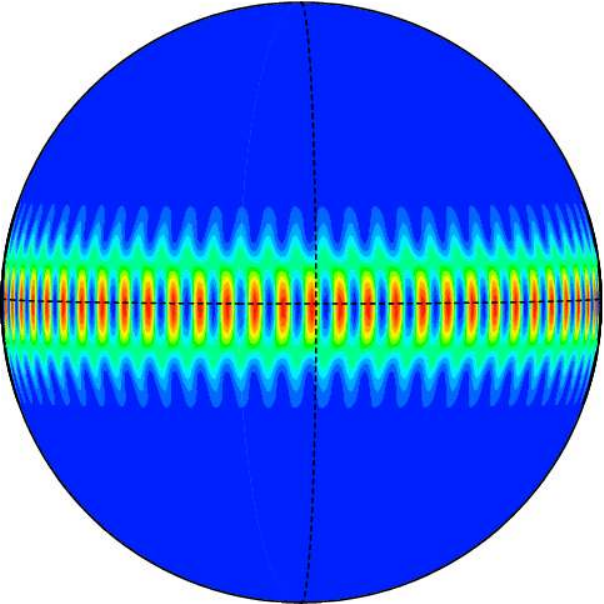}}&
      \resizebox{!}{31mm}{\includegraphics{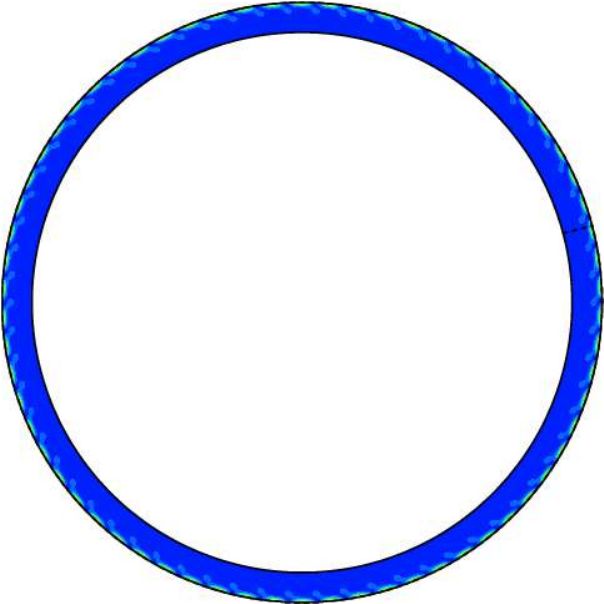}}&
      \resizebox{!}{31mm}{\includegraphics{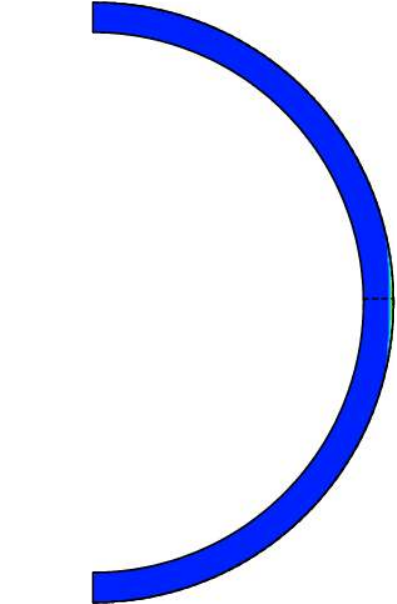}}\\
    \end{tabular}\\
        \vspace{0.4cm}
        \Large{Spiralling Columnar (SC) mode\vspace{0.2cm}}
    \begin{tabular}{cccccc}
      \resizebox{!}{31mm}{\includegraphics{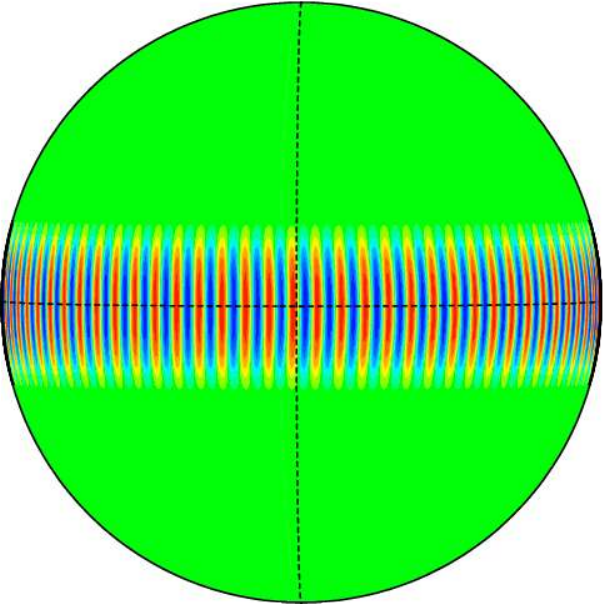}}&
      \resizebox{!}{31mm}{\includegraphics{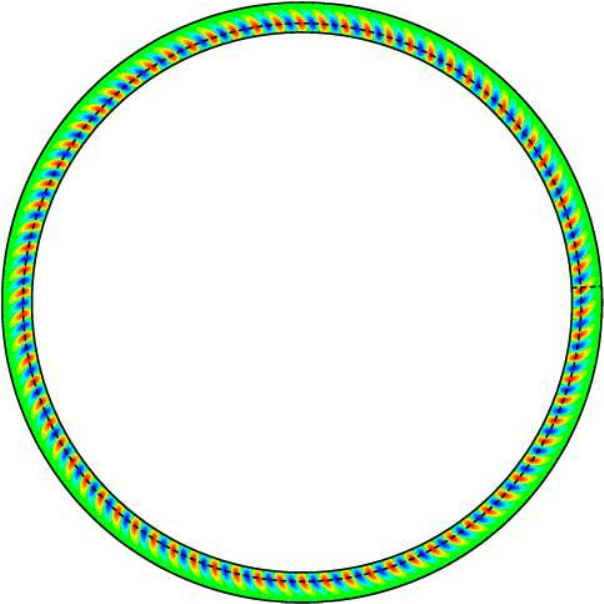}}&
      \resizebox{!}{31mm}{\includegraphics{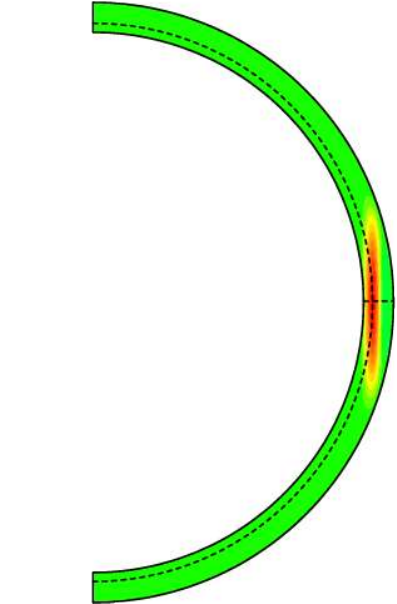}}&
      \resizebox{!}{31mm}{\includegraphics{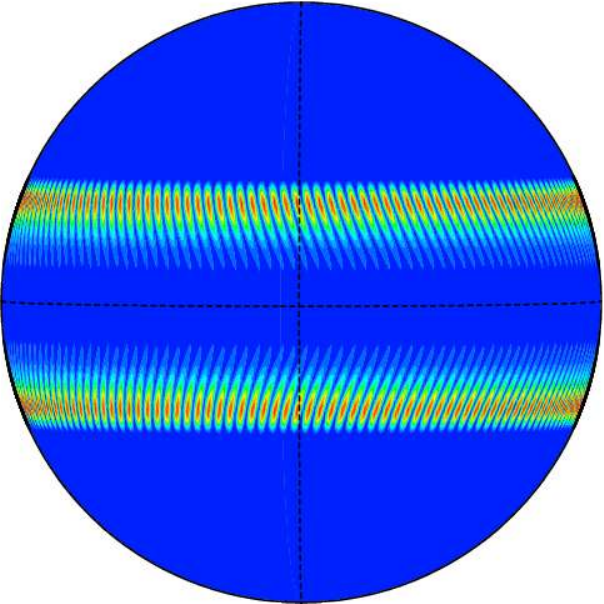}}&
      \resizebox{!}{31mm}{\includegraphics{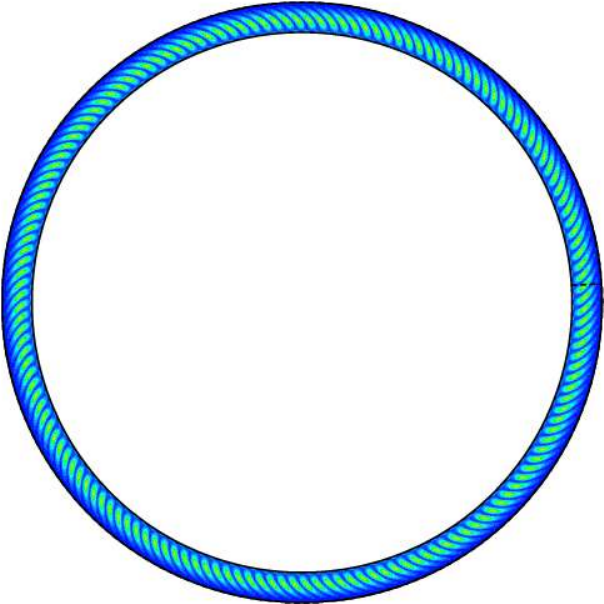}}&
      \resizebox{!}{31mm}{\includegraphics{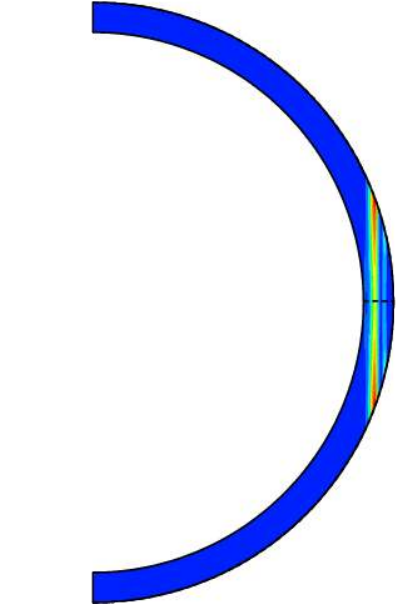}}\\
    \end{tabular}\\
        \vspace{0.4cm}
    \Large{Antisymmetric Polar (AP) mode\vspace{0.2cm}}
    \begin{tabular}{cccccc}
      \resizebox{!}{31mm}{\includegraphics{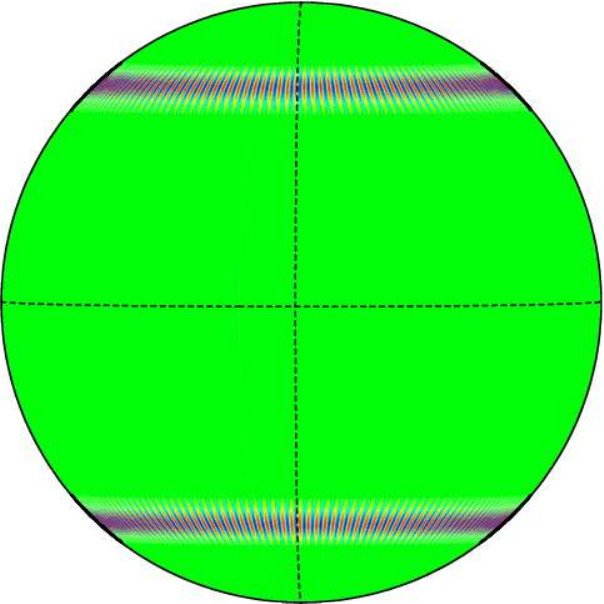}}&
      \resizebox{!}{31mm}{\includegraphics{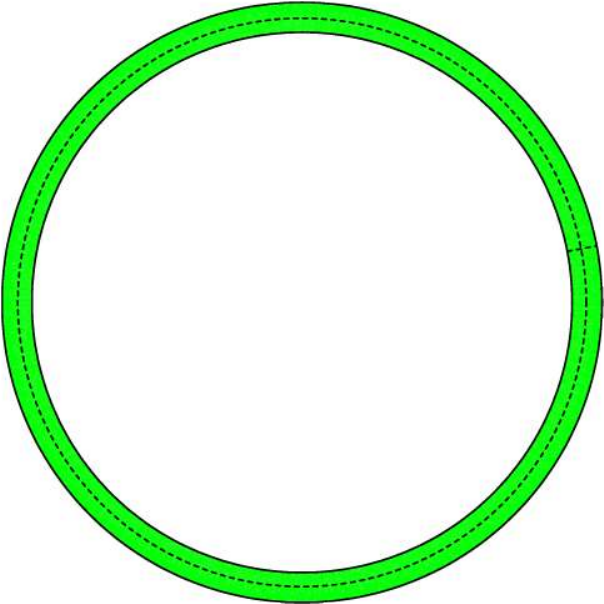}}&
      \resizebox{!}{31mm}{\includegraphics{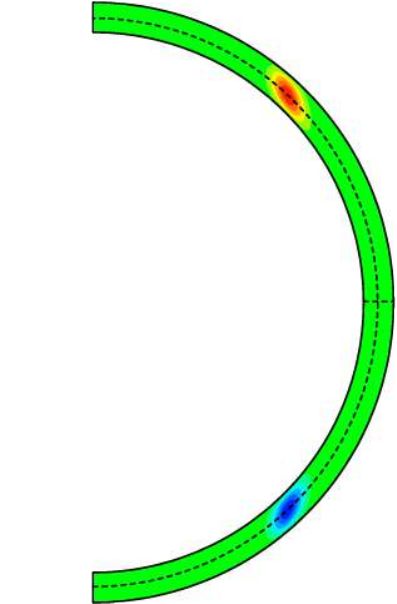}}&
      \resizebox{!}{31mm}{\includegraphics{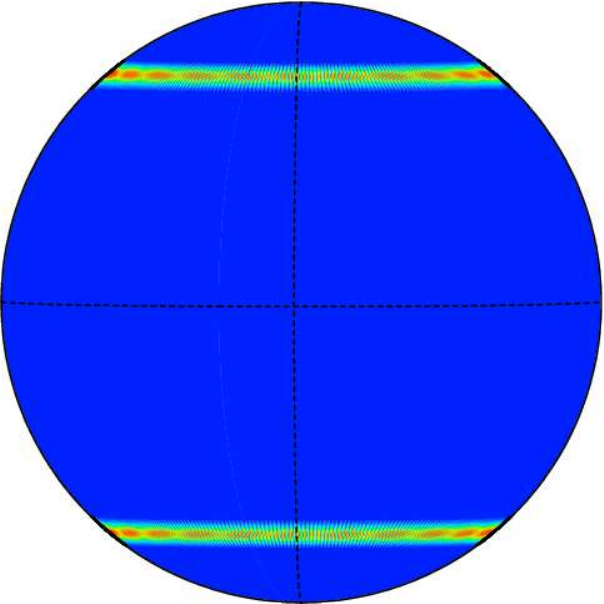}}&
      \resizebox{!}{31mm}{\includegraphics{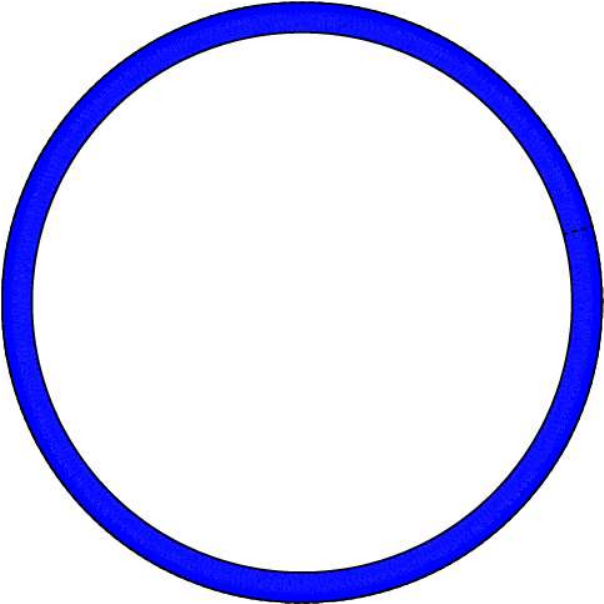}}&
      \resizebox{!}{31mm}{\includegraphics{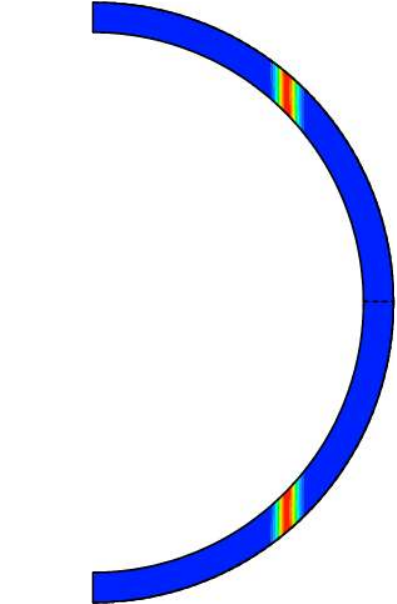}}\\      
    \end{tabular}
    \vspace{0.cm}    
  \end{center}
  \caption{Spherical, equatorial and meridional sections of the
    temperature perturbation $\Theta$ (left) and of the kinetic energy
    density $\ve^2/2$ (right) for the preferred modes of convection at
    $\Pr=10^{-3},\Ta=10^{8}$ with $m_c=18$ (1st row),
    $\Pr=2\times 10^{-3},\Ta=10^{8}$ with $m_c=23$ (2nd row),
    $\Pr=10^{-2},\Ta=10^{8}$ with $m_c=33$ (3rd row),
    $\Pr=10^{-1},\Ta=10^{8}$ with $m_c=78$ (4th row), and
    $\Pr=3\times 10^{-2},\Ta=10^{10}$ with $m_c=156$ (5th
    row). Color scale is not quantitative: $\Theta$: Red/blue means
    hottest/coldest fluid, respectively (green is for
    zero). $\ve^2/2$: Red/blue means most/less energetic fluid,
    respectively. Contour plots of low azimuthal wave number SC,
      EA and AP modes can be found in~\cite{Zha92},~\cite{NGS08}
      and~\cite{GSN08}, respectively.}
\label{fig3} 
\end{figure*}

\begin{figure*}
  \begin{center}
    \begin{tabular}{ccccc}
      AP & SP & EA & SC & AP\\
          &  & &  & \\
      \resizebox{!}{52mm}{\includegraphics{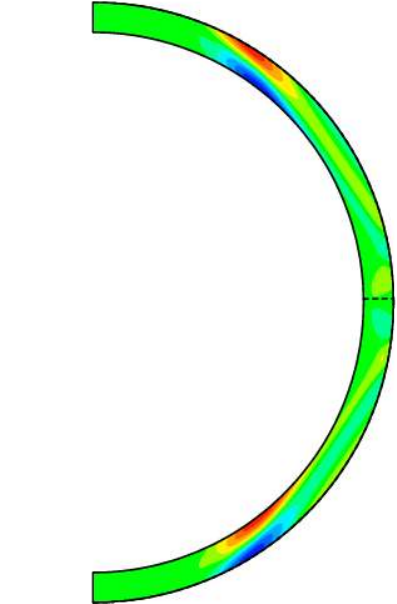}}&
      \resizebox{!}{52mm}{\includegraphics{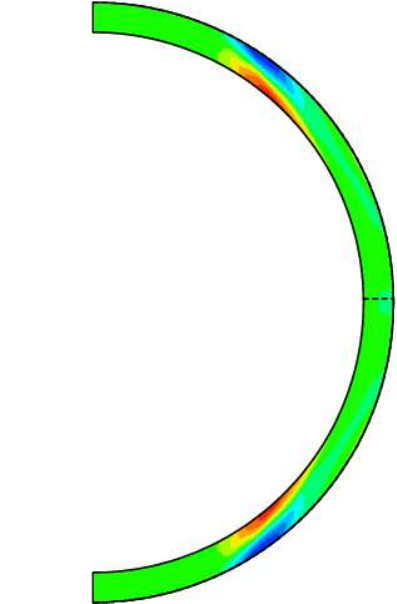}}&
      \resizebox{!}{52mm}{\includegraphics{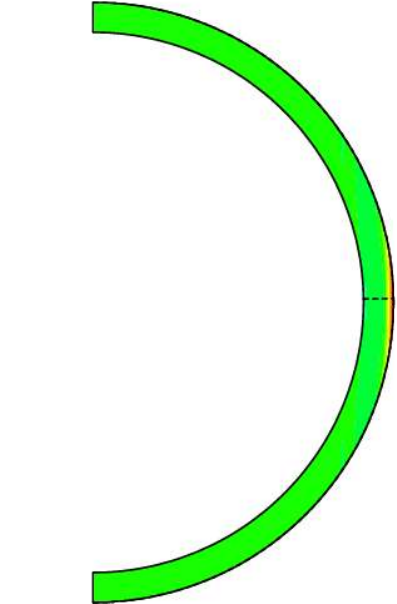}}&
      \resizebox{!}{52mm}{\includegraphics{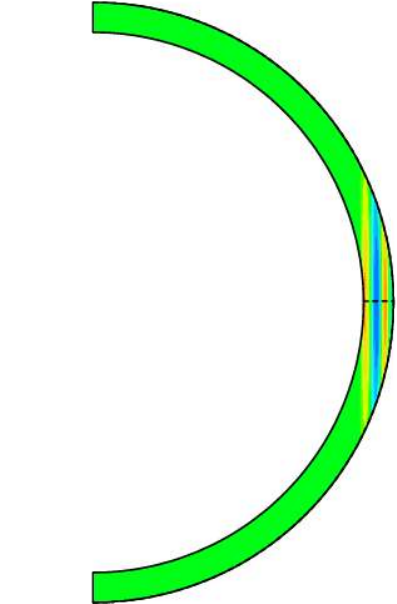}}&
      \resizebox{!}{52mm}{\includegraphics{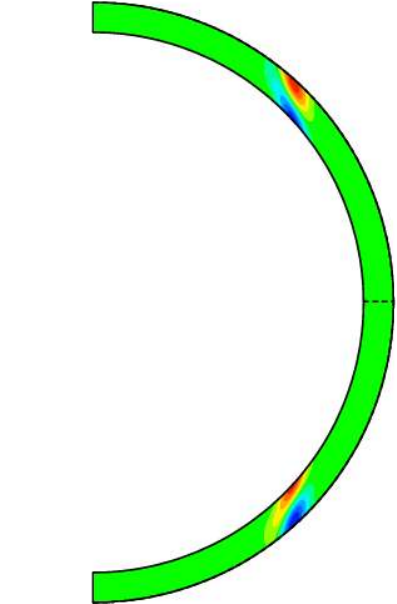}}\\
    \end{tabular}
  \end{center}
  \caption{Meridional sections of the azimuthal velocity $v_{\varphi}$
    for the preferred modes of convection at (from left to right)
    $(\Pr=10^{-3},\Ta=10^{8})$, $(\Pr=2\times 10^{-3},\Ta=10^{8})$),
    $(\Pr=10^{-2},\Ta=10^{8})$, $(\Pr=10^{-1},\Ta=10^{8})$, and
    $(\Pr=3\times 10^{-2},\Ta=10^{10})$. Color scale is not
    quantitative: Red/blue means positive/negative $v_{\varphi}$,
    respectively (green is for zero).  }
\label{fig4}   
\end{figure*}

The flow structure of the spiral columnar modes (SC) is displayed in
the fourth row of contourplots shown in Fig.~\ref{fig3}. The left
group are for the temperature perturbation (from left to right:
spherical, equatorial and meridional sections) and the right group are
for the kinetic energy density $\ve^2/2$. Notice in each section the
black lines showing the position of the other two. The position of the
spherical section is chosen to be close to a relative maximum. In the case
of $\ve^2/2$ it corresponds to the outer boundary. This figure shows
that SC modes are equatorially symmetric and typically elongated in
the axial direction (see the contour lines almost parallel to the
vertical axis in the meridional section of $\ve^2/2$), spiralling
eastward in the azimuthal direction (slightly noticeable in each small
cell of $\ve^2/2$ in the equatorial section) and nearly tangent to the
inner sphere. See also the 4th (from left to right) meridional section
of the azimuthal velocity shown in Fig.~\ref{fig4}.

\begin{figure}[t!]
\begin{center}
\includegraphics[width=0.9\linewidth]{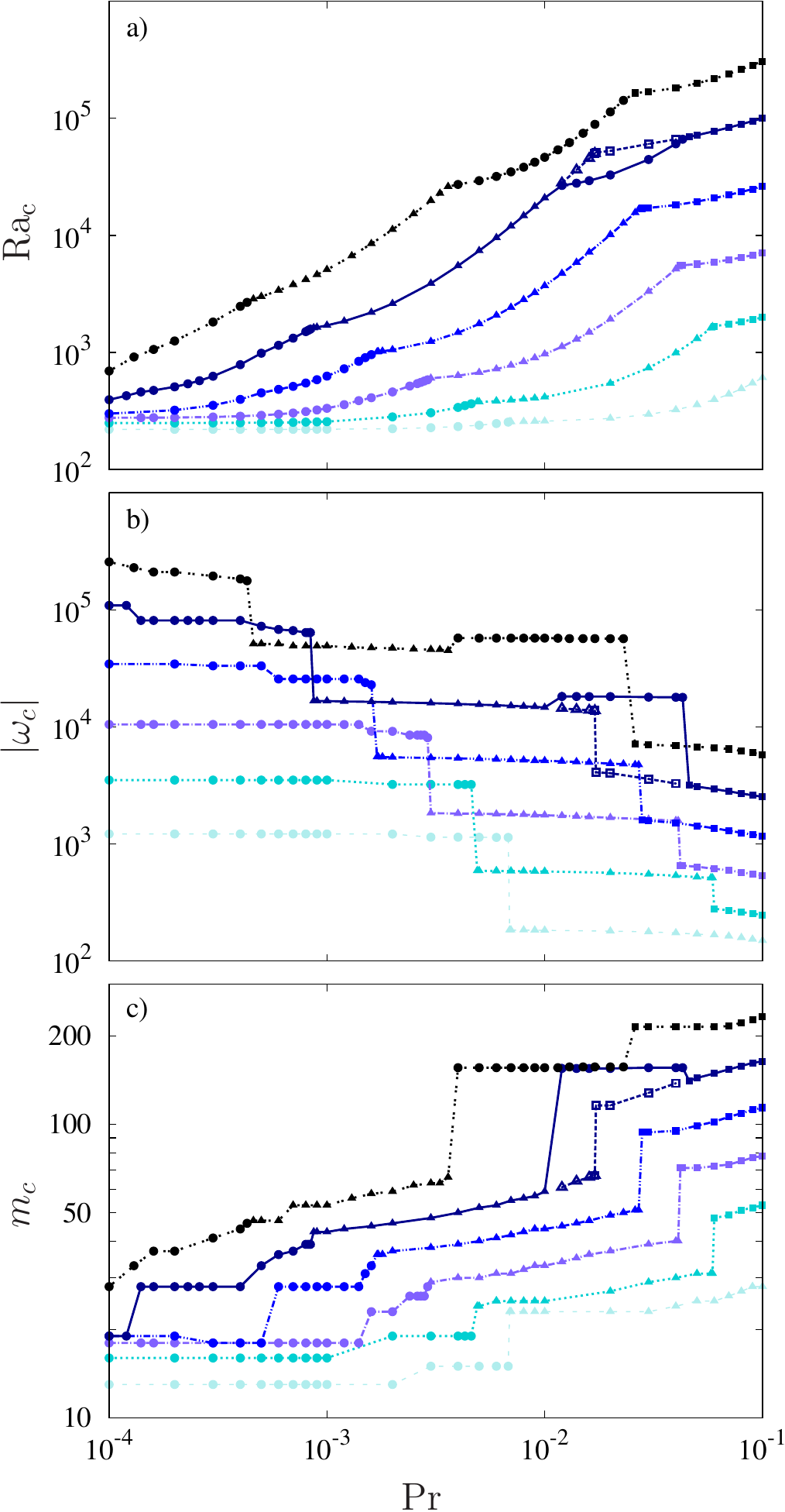}
\end{center}
\caption{(a) Critical Rayleigh numbers $\Rayc$, (b) critical
  drifting frequencies $|\omega_c|$ and (c) critical azimuthal wave
  number $m_c$ versus the Prandtl number $\Pr$ for
  $\Ta=10^{6},10^{7},10^{8},10^{9},10^{10},10^{11}$ (thin dashed,
  dotted, dashed-dotted, dashed-double-dotted, solid and double-dotted
  line style, respectively (blue darkness increases with $\Ta$
  online)). The symbols mean: $\bullet$ AP/SP mode, $\blacktriangle$
  EA mode and $\blacksquare$ SC mode. Full/open points are used for
  dominant/nondominant modes. The thick dashed line marks the
  transition between nondominant EA and SC modes at
  $\Ta=10^{10}$. This figure shows the tendency of AP/SP, EA and SC
  modes to be preferred at smaller $\Pr$ when $\Ta$ increases. For the
  largest $\Ta$, AP/SP modes are preferred at very small but also at
  relatively high $\Pr$.}
\label{fig5} 
\end{figure}

\begin{figure}[t!]
\begin{center}
\includegraphics[width=0.9\linewidth]{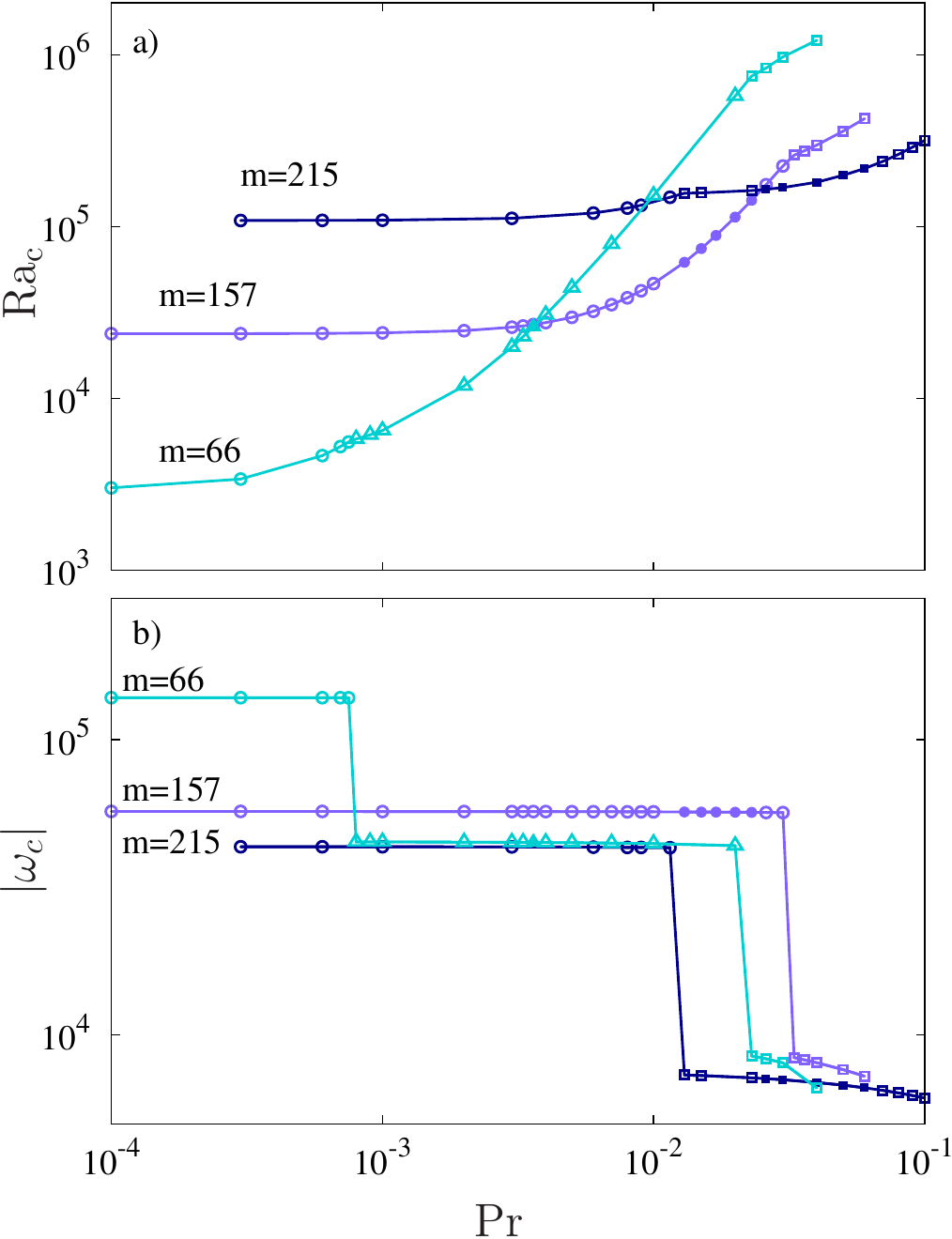}
\end{center}
\caption{(a) Critical Rayleigh numbers $\Rayc$ and (b) critical
  drifting frequencies $|\omega_c|$ versus the Prandtl number $\Pra$
  for $\Ta=10^{11}$ and azimuthal wave numbers $m=66$, $m=157$ and
  $m=215$. The symbols mean: $\bullet$ AP/SP mode, $\blacktriangle$ EA
  mode and $\blacksquare$ SC mode. Full/open points are used for
  dominant/nondominant modes. While for $m=66$ the three types of
  modes are found, for $m=157,215$ only AP/SP or SC are obtained.}
\label{fig6} 
\end{figure}

The situation is quite different for $\Pr=10^{-2}$ and $\Pr=10^{-3}$.
In this case thermo inertial modes become preferred, which can be
either equatorial (EA)~\cite{ZhLi04,NGS08,SGN16} or polar
(P)~\cite{GSN08} depending on $\Ta$. This is shown in Fig.~\ref{fig2}
where the critical parameters are plotted versus $\Ta$. At
$\Pr=10^{-2}$ and $\Ta < 1.25\times 10^{10}$ the instability takes the
form of equatorial waves in which convection is nearly absent in the
bulk of the fluid except for a very thin region near the outer surface
at equatorial latitudes (see 3rd row of Fig.~\ref{fig3} or the 3rd
(from left to right) meridional section of Fig.~\ref{fig4}.). While
for the lowest $\Ta$ the increasing of $\Ray_c$ is quite slow (see
Fig.~\ref{fig2}(a)), at $\Ta \in[10^8,5\times 10^9]$ the power law
fits well with $\Rayc=0.008\Ta^{0.63}$, also found in~\cite{NGS08} for
the EA modes at $\Pr=0.005$ in a non-slip thick internally heated
shell. From around $\Ta=5\times 10^9$ the $\Rayc$ of EA modes follow a
power law with a slightly larger exponent $\propto \Ta^{0.83}$ as found
in~\cite{SGN16} for the EA modes in a stress-free sphere at the same
$\Pr$ and up to $\Ta=10^{14}$. Nevertheless, the only difference
observed between modes following the two different power laws is that
at high $\Ta$ the patterns of $\Theta$ and $v_r$ are significantly
more attached to the outer boundary.

Although at $\Pr=10^{-2}$ the modes are mostly prograde, for the lowest
$\Ta$, retrograde waves have been found (see the jump in
Fig.~\ref{fig2}(b) at $\Ta=4\times 10^4$). This is typical and has
been reported before for EA in thick shells. At $\Ta\in[10^6,5\times
  10^9]$ the variation of $|\omega_c|$ with $\Ta$ can be reasonably
well approximated by $|\omega_c|=0.25 \Ta^{0.48}$ while from
$\Ta=5\times 10^9$ (i.e, when EA become non dominant) it is better to
use $|\omega_c|\propto \Ta^{0.43}$. Both have an exponent a bit larger
than predicted by asymptotic theories but in close agreement with the
case of EA modes on a full sphere~\cite{SGN16}.

The dependence of the critical azimuthal wave number $m_c$ displayed
in Fig.~\ref{fig2}(c) for the EA modes is a little bit lower than
theoretical predictions, following $m_c=3\Ta^{0.13}$ in the interval
$\Ta\in[10^7,10^{10}]$. For lower $\Ta<10^7$, $m_c$ remains nearly
constant while for $\Ta>10^{10}$ (where EA are non dominant) the
exponent seems to increase to $m_c\propto \Ta^{0.18}$. At higher
$\Ta>10^{11}$ the exponent found for EA modes in a full sphere was
0.25.

Surprisingly, and in contrast to what is found in full spheres, at $\Ta
\approx 1.25\times 10^{10}$ EA modes are no longer preferred and
either equatorially symmetric or antisymmetric polar modes, as
in~\cite{GSN08}, are selected. Unlike the latter study with
$\eta=0.4$, this antisymmetric modes have a very large $m_c$ and are
dominant in a significantly larger interval $\Ta \in [1.5\times
  10^{10},10^{12}]$. By comparing our results at $\eta=0.9$ with those
of~\cite{GSN08} at $\eta=0.4$ and those of~\cite{SGN16} at $\eta=0$,
it seems that the thinner the shell, the more probable it is that AP or SP modes
become dominant at high $\Ta$ and moderate $\Pr$. As it will be shown
in the next section, AP or SP modes of relatively small azimuthal wave
number are the most typical at low $\Pr$.

In the range explored $\Ta \in [1.25\times 10^{10},10^{12}]$ the
critical azimuthal wave number of the polar mode remains almost unchanged $m_c\in[155,158]$ (see Fig.~\ref{fig2}(c)) and the frequency 
follows $|\omega_c|\propto \Ta^{0.5}$. The dependence of $\Ray_c$ is
not so clear and no power law can be fitted. When high azimuthal wave
number AP or SP modes become preferred, convection is located in a
narrow band parallel to the equator at very high latitudes, and
$\ve^2/2$ is almost $z$-independent. This is displayed in last row of
Fig.~\ref{fig3} or the rightest meridional section of
Fig.~\ref{fig4}. The latter figure also shows the existence of a
strong shear layer corresponding to a cone which is tangent to the
inner sphere at latitudes close to $30^{\text{o}}$.

At $\Pr=10^{-3}$ inertial prograde modes also become preferred, but in
this case EA modes are dominant at high $\Ta>6.13\times 10^9$ and AP
or SP are dominant at low $\Ta<6.13\times 10^9$ (i.e the situation is
reversed when compared with that at $\Pr=10^{-2}$). The transition can
be clearly seen in the jump (from $|\omega_c|\approx 5\times 10^4$ to
$|\omega_c|\approx 10^4$) of Fig.~\ref{fig2}(b). At such small $\Pr$
the $\Ray_c$ increases slowly with $\Ta$. Only when EA modes become
selected, does the dependence seem to approach to $\propto
\Ta^{0.63}$, and $|\omega_c|$ is roughly $\propto \Ta^{0.5}$ in the
whole $\Ta$ interval. The dependence of $m_c$ is of staircase type but
globally is increasing slowly up to $\Ta\sim 10^{8}$, and beyond this
with a slope $\propto \Ta^{0.18}$. While the patterns of convection of
EA modes at $\Pr=10^{-3}$ are quite similar to those at $\Pr=10^{-2}$,
those of AP or SP modes exhibit differences. The first noticeable
difference is the azimuthal length scale (compare 1st/2nd with last
row in Fig.~\ref{fig3}). A second difference is that at $\Pr=10^{-2}$
the $z$-dependence of $\ve^2/2$ is enhanced. This is not surprising
because these polar modes are dominant at larger $\Ta$. In addition,
at $\Pr=10^{-3}$ the modes are weakly multicellular (see spherical and
meridional sections of $\Theta$) with convection present at larger
latitudes and with shear layers extending from polar latitudes to
close to the equator (see 1st/2nd plot (from left to right) of
Fig.~\ref{fig4}). Notice that there are no fundamental differences in
the patterns of the AP and SP modes. For such a thin shell, any link
between the high latitudes on both hemispheres practically disappears,
and an AP mode can be obtained by azimuthally shifting an SP mode on
one of the hemispheres.

Finally, at $\Pr=10^{-4}$ only SP or AP modes like those at
$\Pr=10^{-3}$ are preferred up to $\Ta=10^{12}$. Up to $\Ta\approx
10^8$ there is almost no difference in the critical parameters between
$\Pr=10^{-3}$ or $\Pr=10^{-4}$ (see Fig.~\ref{fig1}). In addition,
$\Ray_c$ starts to increase very slowly from $\Ta>10^{10}$ and no
power law can be fit. Polar modes drift with very large (up to $10^6$)
frequencies, that follow $|\omega_c|=1.5\Ta^{0.48}$. For such small
$\Pr=10^{-4}$, the modes can be prograde but also retrogade as it
happens for $\Ta=10^{9},10^{10}$ and $m_c=19$. The azimuthal wave
number $m_c$ also starts to increase from $\Ta>10^{10}$ with
$m_c=0.3\Ta^{0.18}$, i.e, with the same slope as for $\Pr=10^{-3}$
from $\Ta>10^8$ ((see Fig.~\ref{fig1}(c)).  Then, for the polar modes
the large $\Ta$ limit (where $m_c$ starts to increase) seems to be
reached at $\Pr \Ta^{1/2}\sim 10$.

\section{Prandtl number dependence}
\label{sec:Pr}

According to the previous section, different types of modes may be
preferred depending on $\Pr$. The aim of this section is to find the
transitions between these modes in the parameter space $(\Ta,\Pr)$. To
that end, further exploration by varying $\Pr$ is needed.

Figure~\ref{fig5} shows $\Ray_c$, $|\omega_c|$ and $m_c$ as a function of
$\Pr\in[10^{-4},10^{-1}]$ for a set of 6 Taylor numbers from
$\Ta=10^{6}$ to $\Ta=10^{11}$. In this figure the transition between
spiralling, equatorial and polar modes can be identified by the cusps
in the $\Ray_c$ curves (Fig.~\ref{fig5}(a)) and jumps in the
$|\omega_c|$ (Fig.~\ref{fig5}(b)) or $m_c$ (Fig.~\ref{fig5}(c))
curves. By increasing $\Pr$ from $10^{-4}$ there exist first one or
several transitions among SP and AP modes for all $\Ta$ explored (some
of them can be retrograde at $\Pr\sim 10^{-4}$). This is clearly
noticeable from Fig.~\ref{fig5}(c) in the jumps in each of the $m_c$
curves at the left of the plot. The convection patterns of these polar
modes were described in previous section and are shown in the 1st (AP)
and 2nd (SP) rows of Fig.~\ref{fig3} and in the 1st (AP) and 2nd (SP)
plot of Fig.~\ref{fig4}. For transitions between polar modes (either
symmetric or antisymmetric) the jumps in $|\omega_c|$ are very small as are
 the cusps in $\Ray_c$.

The second transition, observed at a critical $\Pr$ number which tends
to decrease with increasing $\Ta$, is between polar and
equatorial modes. At the transition, $|\omega_c|$ decreases sharply by
roughly one order of magnitude and $\Ray_c$ increases, with smaller
slope. There exists also a jump in $m_c$, and the jump tends to
decrease with $\Ta$.

For $\Ta\le 10^9$ and larger $\Pr$ a third transition between EA and
SC modes is found. As for the transition between SP/AP and EA modes,
the critical $\Pr$ of this transition tends to decrease with
increasing $\Ta$ although at a slightly slower rate. This
transition is also characterised by a decrease of $|\omega_c|$ but is
not so pronounced. When the new modes become preferred $\Ray_c$
increases at a lower rate, and they have a very small azimuthal length
scale (seen as a big jump in $m_c$).

As mentioned in previous section, we have found very high azimuthal
wave number polar modes preferred at moderate $\Pr$ and large
$\Ta$. They are clearly dominant at $\Ta=10^{10},10^{11}$ around
$\Pr=10^{-2}$ (see right top part of Fig.~\ref{fig5}) and their region
of stability tends to increase with $\Ta$. In the curve for
$\Ta=10^{10}$ the transition between the non dominant EA and SC modes
is shown (the open symbols on the right part of that curve). To
visualise how these polar modes become dominant $\Ray_c$ and
$|\omega_c|$ are plotted in Fig.~\ref{fig6} versus $\Pr$ for the
azimuthal wave numbers $m=66,157,215$ at $\Ta=10^{11}$. While for
$\Pr\in[10^{-4},10^{-1}]$ the $m=66$ mode can be polar, equatorial or
spiralling (notice the two transitions in jumps of $|\omega_c|$), the
$m=157,215$ modes only show a transition between AP/SP and SC
modes. When decreasing $\Pr$ beyond the transition, $\Ray_c$ starts to
decrease faster. The fact that the transition occurs at larger $\Pr$
for the $m=157$ polar mode allows it to be dominant.

We confirm the trend occurring in thicker shells. The smaller the
$\Pr$ the smaller $\Ray_c$ and $m_c$ and the larger $|\omega_c|$. In
the $\Pr$ interval explored, due to the multiple transitions, $\Ray_c$
cannot be fit by simple power laws. The drifting frequencies remain
nearly constant for each type of mode, with the SC modes having a
slightly steeper decrease. The dependence of $m_c$ with $\Pr$ can be
approximated as roughly $\propto \Pr^{0.14}$ when $\Ta>10^6$ for the
EA modes and $\propto \Pr^{0.2}$ for the SC modes and the larger
$\Pr$. This contrast with the results on the full sphere~\cite{SGN16}
where an exponent of $0.54$ was found for the EA modes.

\section{Transitions in the $(\Ta,\Pr)$ parameter space}
\label{sec:tran}

\begin{figure}[t!]
\begin{center}
\includegraphics[width=0.9\linewidth]{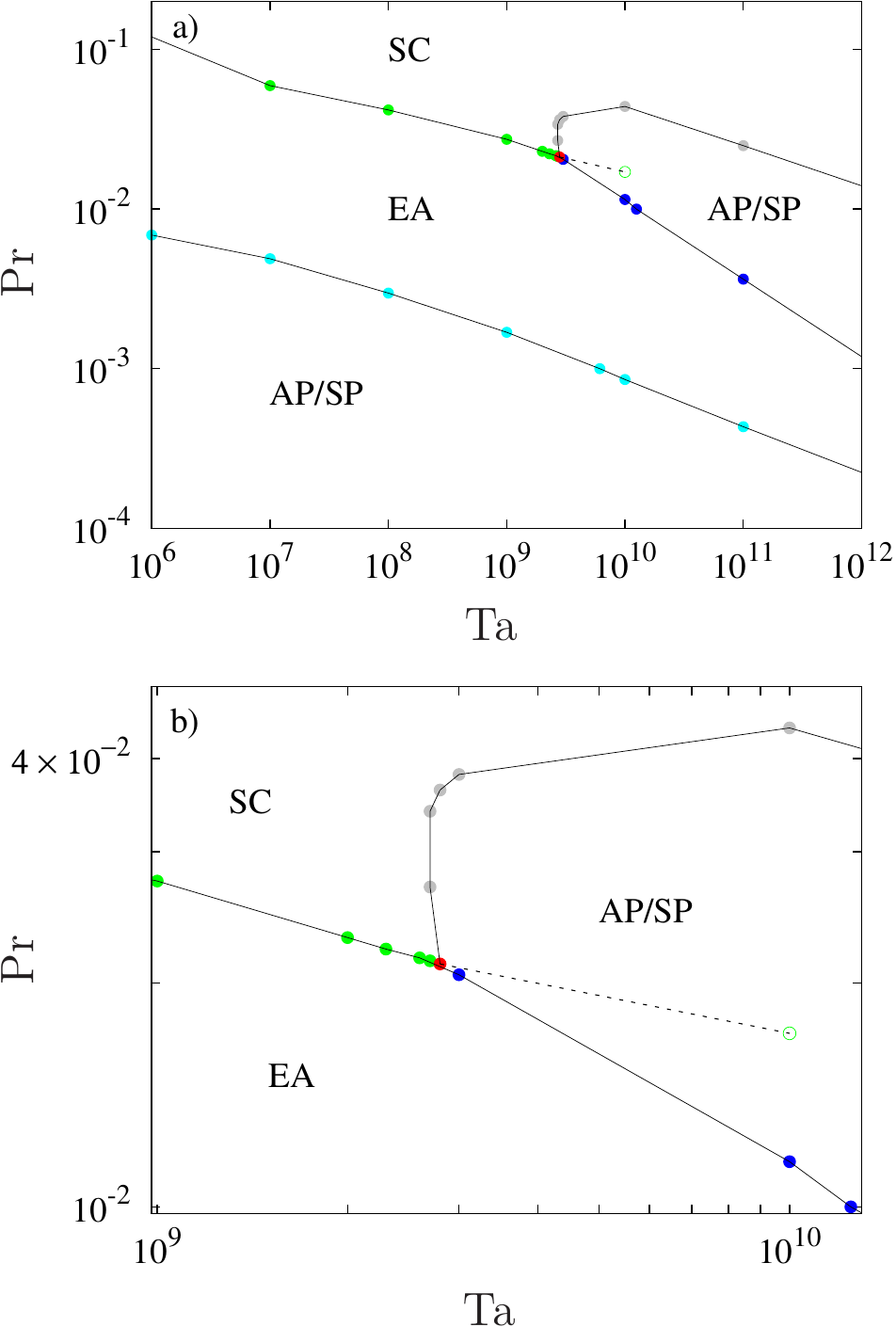}
\end{center}
\caption{(a) Transitions between the different types of preferred
  modes (the points are obtained from Fig.~\ref{fig2} and
  Fig.~\ref{fig5} by means of linear interpolation). (b) Detail of (a)
  showing the triple point. The dashed line and the open circle mark
  the transition between nondominant EA and SC modes shown in
  Fig.~\ref{fig5}.}
\label{fig7} 
\end{figure}

From Fig.~\ref{fig2} and Fig.~\ref{fig5} and some additional
computations, not shown in those figures, the locations of the
different transitions in the $(\Ta,\Pr)$ parameter space have been
obtained by linear interpolation, and the regions of stability of the
different modes have also been roughly established. This is shown in
Fig.~\ref{fig7} where the transitions from AP/SP to EA modes, from EA
to SC modes, from EA to high $m$ AP/SP modes, and from high $m$ AP/SP
to SC modes, correspond to the $4$ different curves containing cyan,
green, blue and gray points, respectively. The general trend of this
figure is that polar modes are preferred at very low $\Pr$, equatorial
modes at moderate $\Pr$ and spiral modes at larger $\Pr$, but polar
modes with high azimuthal wave number can also be preferred at
moderate $\Pr$. As $\Ta$ is increased the position of the transitions
tend to be located at lower $\Pr$. For instance we have obtained
$\Pr_{\text{P/EA}}=0.68\Ta^{-0.29}$ (cyan points) and
$\Pr_{\text{EA/SC}}=1.67\Ta^{-0.2}$ (green points), both power laws
for $\Ta>10^8$. At approximately $(\Ta_3,\Pr_3)=(2.8\times
10^{9},2.12\times 10^{-2})$ there is a transition between EA, AP and
SC modes, giving rise to a triple Hopf bifurcation. At this point an
equatorial, a polar and a spiralling mode have equal critical
Rayleigh numbers within $1\%$ of precision (Table~\ref{tab1}), i.e, of
the same order of the spatial truncation errors. To the best knowledge of
the authors this is the first time a triple point has been reported in the
context of thermal convection in rotating spherical geometry. Choosing
parameters near $(\Ta_3,\Pr_3)$ will give rise to a very rich variety
of travelling and modulated waves with very different azimuthal wave
numbers, convective patterns and energy balances.

The region of stability of the high $m$ polar modes seems to widen
with $\Ta$ making the region of stability of equatorial
modes thinner. By extrapolating the EA/SC transition curve to larger
$\Ta>10^{12}$ (by continuing the dashed line through the empty circle
and so on in Fig.~\ref{fig7}) one might expect the EA/SC transition
curve to appear again, arising at another triple point. This is likely to
happen because the slope of the P/SC transition curve seems to be
larger than that of the EA/SC modes. A similar argument can be applied
to conclude that the region of equatorial modes confined between the
two regions of polar modes will probably disappear, so that at some
fixed $\Ta>10^{12}$ there will exist only 3 types of modes when $\Pr$
is varied, namely (from low to large $\Pr$) AP/SP, EA and SC modes. An
interesting question then arises: could it be that for some $\Ta>10^{12}$
 high wave number AP/SP modes become preferred again between
EA and SC modes?

\begin{table}[ht] 
  \begin{center}
    \begin{tabular}{lccccccccccc}
\vspace{0.1cm}            
& Type & & $m_c$      & $\Ray_c$      & $\omega_c$     \\
\hline\\
&EA&  &$57$      & $2.5564 \times 10^4$      & $-7.6968\times 10^3$     \\
\\
&SC&  &$92$      & $2.5662 \times 10^4$      & $-2.6474\times 10^3$     \\
\\
&AP&  &$154$     & $2.5699 \times 10^4$     & $-9.6376\times 10^4$     \\
\\
\hline
  \end{tabular}
  \caption{Critical parameters for equatorial, spiralling and polar
    modes of convection at $\Ta=2.8\times 10^9$ and $\Pr=2.12\times
    10^{-2}$. The differences between all $\Ray_c$ are less than $1\%$
    (assumed for spatial resolution errors). Then, at this $(\Ta,\Pr)$
    an AP, EA and SC mode are preferred at the same time. }
  \label{tab1}
  \end{center}
\end{table}

\section{Application to convection in stars and stellar oceans}
\label{sec:star}

For the sake of curiosity, the critical parameters of the onset of
convection for three different type of astrophysical applications will
be estimated in the following with the help of the power law scalings
obtained in previous sections. They are:
\begin{itemize}
\item Spiralling modes: $\Rayc \sim 0.2 \Ta^{0.57}$, $|\omega_c|\sim
  \Ta^{0.34}$  and $m_c\sim 4 \Ta^{0.16}$.
\item Equatorial modes: $\Rayc\sim 0.008\Ta^{0.63}$, $|\omega_c|\sim
  0.25\Ta^{0.48}$ and $m_c \sim 3\Ta^{0.13}$.
\item Polar modes: $|\omega_c|\sim 1.5\Ta^{0.48}$ and $m_c \sim
  0.3\Ta^{0.18}$ for $\Pr \Ta^{1/2} >10$.
\end{itemize}

In addition, the type of instability (SC,EA or P) can be predicted
using $\Pr_{\text{P/EA}}=0.68\Ta^{-0.29}$ and
$\Pr_{\text{EA/SC}}=1.67\Ta^{-0.2}$ obtained in
\S\ \ref{sec:tran}. The Taylor number, $\Ta=\Omega^2 d^4/\nu^2$, and
Prandtl number, $\Pr=\nu/\kappa$, come from values of physical
properties (see Table~\ref{tab2}) obtained in previous studies,
details of which are given below. We also estimate the period of the
waves from $P=(2\pi m_c/|\omega_c|)d^2/\nu$ in seconds. Recall that
negative frequencies give positive drifting (phase) velocities
$c=-\omega_c/m_c$ and the waves drift in the prograde direction.

At this point, it is worth mentioning the applicability of these
results to stellar convection. In our Boussinesq model, density
variations are neglected, except in the buoyancy term. This is far from
stellar conditions, where the scale height of density
variations~\cite{SpVe60} can be three orders of magnitude smaller than
the gap width of the convective layer. However, this simplification
can be justified when the focus is to study the basic but relevant
mechanisms of rotation, buoyancy and thin spherical geometry, and their
interconnections.

In addition, it is often stated that qualitative dynamical properties
of Boussinesq flows are inherited by compressible flows at the same
parameter conditions, provided the velocities remain subsonic.  For
instance, quite similar large scale two vortex turbulent structures
have been found in~\cite{ChGl05} in a rectangular box both with and
without strong stratification, in both cases with kinetic energy
scaling as $k^{-3}$ ($k$ is the Kolmogorov wave number). In the
specific case of rotating spherical shells, the linear theory of
anelastic convection was formulated in~\cite{JKM09}. They found that
with stratification, convection tends to be moved to near the outer
shell with larger $\Rayc$, $m_c$ and $|\omega_c|$ for
$\Pr=0.1,1,10$. Specifically, with the strongest stratification,
$\rho_{r_i}/\rho_{r_o}\approx 150$, the critical parameters increase,
with respect to the Boussinesq case, by around an order of magnitude
in the worse case. With these considerations, the linear theory of
Boussinesq convection may be used to obtain reasonable bounds for the
critical parameters for compressible flows in a regime where the
compressible formulation is still numerically unfeasible. Further
support for this result is given in~\cite{CJM15}. The same scaling
laws for the critical parameters, in both Boussinesq and compressible
linear analysis, are valid in the limit of large rotation in a
rotating plane layer geometry with very low $\Pr$. As the authors
of~\cite{CJM15} stated in their study, further research at the very
low $\Pr$ number regime is needed. Recently~\cite{WoBu16}, also in a
rotating plane layer geometry, found restrictions on the Boussinesq
approximations at very low $\Pr$. In this study, asymptotic scalings
for the Boussinesq and compressible onset of convection were compared
with numerical computations of the compressible case. For a perfect
gas, with moderate ratio between the domain and the temperature scale
heights, differences of less than one order of magnitude in $\Ray_c$
were found between the two approximations, at $\Pr=10^{-4}$.

Although convection in stars is believed to be fully
turbulent~\cite{BrTo95,BNS00}, the analysis of the onset of convection
is worthwhile because patterns and characteristics at the onset may be
able to persist even in strongly nonlinear regimes. This was argued to
happen in~\cite{LaAu11} even in the case of nondominant inertial and
axisymmetric modes. A previous study~\cite{GBNS14} also suggests that
the mean period of a fully developed turbulent solution does not
change drastically from that of the onset of convection. In that study
at $\Pr=0.1$, the frequency spectrum of time series from turbulent
solutions that are 100 times supercritical have a maximum peak at a
frequency roughly 3 times larger than the corresponding value at the
onset of convection.

\begin{table*}[t]
  \begin{center}
  \begin{tabular}{llcccc}
  \hline\\
&  Property                &   Sun$^{(1)}$       &  Accreting white    & Accreting neutron  & Molecular region \\
&               &                               &  dwarf ocean$^{(2)}$ &  star ocean$^{(3)}$ & Jupiter$^{(4)}$ \\[3pt]
  \hline \\
& $\kappa$ (cm$^2$s$^{-1}$) & $10^{7}-10^{9}$     & $10^{1}-10^{5}$       &  $10^6$            & $10^{-5}$\\  
& $\nu$ (cm$^2$s$^{-1}$)    & $10^{0}-10^{4}$     & $10^{-4}-10^{-2}$     &  $10^6$            & $10^{-6}$\\
& $d$ (cm)                 & $10^{10}$           & $10^{8}$            &  $10^{3}-10^{4}$     & $10^{8}$\\
& $\Omega$ (s$^{-1}$)       & $2.6\times 10^{-6}$ & $7\times 10^{-5}$   &  $10^{2}$            & $2\times 10^{-4}$\\
  \\
  \hline
\end{tabular}
\end{center}                                                                                                                         
\caption{Physical parameters of three example astrophysical convection
  scenarios. Jupiter is included for comparison purposes. Thermal
  diffusivity $\kappa$, kinematic viscosity $\nu$, layer width $d$ and
  rotation rate $\Omega$.  Values are taken: (1) from~\cite{BNS00} and the references
  therein. (2) from~\cite{NaPe84,GMBMM15,IGKL17}. (3)
  from~\cite{Yakovlev80,NaPe84,Wat12}. (4) from~\cite{BDM04,SKF11}}
\label{tab2}
\end{table*}


For illustration, we shall now consider a selection of different
astrophysical scenarios where convection is important, to assess how
the results of our study would apply.  Shell or envelope convection
can arise in many different astrophysical scenarios.  Low mass main
sequence stars burning hydrogen to helium in their cores, for example,
are expected to have convective zones in their envelopes. In Section
\ref{Sun} we consider this scenario and look at the parameter space
appropriate to the Sun.  Shell convection can also occur during post
main sequence evolution, via shell burning or as elements heavier than
hydrogen are burned.  This may occur on the Red Giant Branch,
Horizontal Branch, and Asympotic Giant Branch (see for example
\cite{Herwig06,Meakin07,Mocak09,Kippenhahn13}).  We do not examine
these scenarios in this paper, but note them for reference.  White
dwarfs can also develop convection in various ways.  Isolated white
dwarfs develop convective zones as they cool, and we examine this
scenario in Section \ref{WD}.  Accreting white dwarfs and neutron
stars can also develop convective zones as accreted material from a
companion star undergoes thermonuclear burning.  When this burning is
unstable, this can manifest in bright transient bursts: classical
novae on accreting white dwarfs, and Type I X-ray bursts on accreting
neutron stars.  We do not study the convective zones of accreting
white dwarfs further in this paper, but refer the interested reader
to, for example, \cite{Shankar92,Glasner95,Kercek99,Starrfield16}.
However in Section \ref{NS} we consider the parameter space
appropriate for accreting neutron star oceans.

\subsection{Sun}
\label{Sun}

For convection occurring in the Sun (\cite{Mie00,Pat10} provide good
reviews focusing on the role of rotation), kinematic viscosities range
from $\nu=1$ cm$^2$s$^{-1}$ near the surface to $\nu=10^4$
cm$^2$s$^{-1}$ in the deeper layers, due to radiative viscosity
(see~\cite{BNS00} and the references therein). This gives rise to very
low Prandtl numbers $\Pr\in[10^{-9},10^{-3}]$ and, depending on the
size $d$ of the convective layer, to quite large Rayleigh numbers
$\Ray\in[10^{18},10^{24}]$ (see Sec. 2.4 of~\cite{Gil00}, Sec. 2.3
of~\cite{BGHNP15} or~\cite{HGS16}). By assuming a size $d=10^{10}$ cm
($\eta>0.8$)~\cite{BNS00}, one expects $\Ta\in[10^{21},10^{29}]$. Our
predictions are shown in Table~\ref{tab3}. According to this table,
given the low Prandtl number of the Sun, the most probable type of
mode is equatorial. This type of Boussinesq solar convection was first
studied in~\cite{Bus70b} by considering only the radial dependence for
small amplitude convection at small $\Ta$. Recent studies on solar
convection in the context of rotating spherical shells incorporate
more complex physics by using the anelastic
approximation~\cite{Gla84}. More realistic parameters ($\Ta=10^6$ and
$\eta\approx 0.7$) have been considered in~\cite{METCGG00}, and weak
lateral entropy variations at the inner boundary have been imposed
in~\cite{MBT06}. Nevertheless, they are restricted to moderate
$\Pr\sim 0.1$. According to our results, and assuming the lowest
$\Pr<10^{-6},$ it is also possible that polar modes become preferred,
or at least become relevant when nonlinearities are included.

At this point it is
interesting to see which time scales come from our results at the
above estimated range of $\Ta$.  Considering an equatorial mode, a long
timescale, $P\sim 10^{3}-10^{4}$ years, is obtained. In the case of
polar modes, the timescale is a little bit shorter, at $10^2-10^4$
years. These orders of magnitude estimates  are not dissimilar to the long
term variability timescales of the Sun, the Gleissber and Suess (also
named the de Vries) cycles of $50-140$ years and $170-260$ yrs,
respectively~\cite{ONKJ02}, an unnamed $500$ and $1000$ yrs cycle and
the Hallstatt cycle ($2300$ years)~\cite{KHPMS12}, or the evidence of
millenial periods of $6000$ yrs~\cite{XaBu09} and $9500$
years~\cite{San16} suggested recently. Notice the robustness of the
predictions: although $\Ta$ spans 8 orders of magnitude, the periods
obtained vary by only 2 orders of magnitude.
Many of the principal features of the well-known 11-year Schwabe
period can basically be explained in terms of dynamo
theory~\cite{Par55,CCJ07}. However, its origin is still not well
understood because the fields are strongly influenced by rotation and
turbulent convection~\cite{SBCBN17}. Moreover, the long-term
solar-activity variation described above imposes several constraints
on current solar dynamo models (either deterministic or including
chaotic drivers). See the very recent review~\cite{Uso17} Sec.~4.4 for
further details. Because convection is believed to be the main driver
of natural dynamos~\cite{DoSo07}, and the different dynamo branches
first bifurcate from purely convective flows, the time scales
predicted from our results may be present in dynamos bifurcated from
convective flows at similar range of parameters. Simulating such
self-excited dynamos at the parameter regime covered in our study is
still numerically unattainable.

\begin{table*}[]
  \begin{center}
  \begin{tabular}{lllcccc}
  \hline
& Type of mode  &  Property       &   Sun    & Accreting white  & Accreting neutron  \\
&   &       &       &  dwarf ocean  &  star ocean \\[3pt]
  \hline                                                                                              \\
  &               & $\Ta^{}$         &  $10^{21}-10^{29}$  & $10^{28}-10^{32}$ &   $10^{4}-10^{8}$       \\
&               & $\Pr^{}$         &  $10^{-9}-10^{-3}$  & $10^{-9}-10^{-3}$ &   $10^{1}-10^{2}$       \\  
&               & $\Pr_{\text{P/EA}}$ & $10^{-9}-10^{-6}$ & $10^{-10}-10^{-8}$ &   $10^{-3}-10^{-1}$       \\
&               & $\Pr_{\text{EA/SC}}$ & $10^{-6}-10^{-4}$  & $10^{-6}-10^{-5}$ &  $10^{-2}-10^{0}$        \\
\\
\hline
\\
&               & $\Rayc$         &  $10^{11}-10^{16}$  & $10^{15}-10^{17}$ &   $10^{3}-10^{6}$       \\
&               & $|\omega_c|$      & $10^{7}-10^{10}$     & $10^{9}-10^{11}$ &   $10^{0}-10^{1}$       \\
& SP            & $m_c$           &  $10^{4}-10^{5}$    & $10^{5}-10^{6}$  &   $10^{0}-10^{1}$       \\
&               & $P$)(years)      &  $10^{6}-10^{9}$    & $10^{7}-10^{8}$  &   $10^{-8}-10^{-6}$       \\
\\
\hline
\\
&               & $\Rayc$         &  $10^{11}-10^{16}$    & $10^{15}-10^{18}$ &        \\
&               & $|\omega_c|$      & $10^{9}-10^{13}$     & $10^{13}-10^{14}$ &        \\
& EA           & $m_c$           &   $10^{3}-10^{4}$          & $10^{4}$  &        \\
&               & $P$ (years)       &  $10^{3}-10^{4}$      & $10^{2}-10^{3}$ &         \\
\\
\hline
\\
&               & $\Rayc$         &                    &                    &         \\
&               & $|\omega_c|$      &  $10^{10}-10^{14}$     &  $10^{13}-10^{15}$   &          \\
& P             & $m_c$           &   $10^{3}-10^{4}$     &  $10^{4}-10^{5}$     &          \\
&               & $P$ (years)       &    $10^{2}-10^{4}$     & $10^{2}-10^{3}$     &          \\
\\
  \hline
\end{tabular}
\end{center}                                                                                                                         
\caption{Estimation of the critical parameters and a typical time
  scale $P=(2\pi m_c/|\omega_c|)d^2/\nu$ in seconds (this is the
  timescale associated with the drifting phase velocity as computed in
  the rotating frame) for the Sun, a white dwarf and a neutron star
  depending on the type of mode, spiralling (SC), equatorial (EA) or
  polar (P). In the case of neutron stars only SC modes are considered
  since they are the only ones that can be preferred for the range of
  parameters. For the Sun and the white dwarf scenario considered, all
  of them could be preferred depending on appropriate particular
  choice of $\Pr$. This is indicated by the critical $\Pr$ at the P/EA
  and EA/SC transitions, which are also shown.}
\label{tab3}
\end{table*}

\subsection{Cooling White Dwarfs}
\label{WD}

%
%


White dwarfs are highly degenerate, except in a thin layer close to
the surface.  The idea that cooling white dwarfs may have a convective
mantle is a long-established one (see for example
\cite{Schatzman58,Fontaine76, Tremblay13}).  Very recently,
~\cite{IGKL17} suggested that associated dynamo action might explain
the magnetic field observed in isolated white dwarfs. The authors
focused on a heavy ($1.0M_{\odot}$) white dwarf because magnetic white
dwarfs are typically observed to be more massive.  Here we consider a
cooling white dwarf with properties similar to the one studied in that
paper, noting however that convective envelopes of white dwarfs could
occupy a much wider range of parameter space.

Following~\cite{NaPe84} a kinematic viscosity of $\nu=3.13\times
10^{-2}$ cm$^2$s$^{-1}$ was used in~\cite{IGKL17}. We consider the
range $\nu=10^{-4}-10^{-2}$ cm$^2$s$^{-1}$ of possible viscosities,
take $d=10^8$cm as the size of the convective layer (see Figure 1 of
\cite{IGKL17}), and a total stellar radius roughly $R=5\times 10^8$cm
for our estimations. These values give rise to $\eta\approx
0.8$. However, as the white dwarf cools, the size of the convective
layer decreases towards zero, giving rise larger values of
$\eta$. Although periods of white dwarfs range from hours to days or
longer~\cite{Kaw04}, we assume a period of one day. Results are shown
in table~\ref{tab3}. As in the case of the Sun, the expected low
Prandtl number makes equatorial modes more feasible, but polar modes
are also possible. Convection sets in at a critical Rayleigh number
$10^{15}-10^{18}$, with drifting periods $\sim 10^2-10^3$ years.
Further work would be required to investigate the potential
consequences.


\subsection{Accreting Neutron Stars}
\label{NS}

The case of convection in very thin oceans of accreting neutron stars
covers quite a different region in Ta, Pr parameter space compared to
the Sun and white dwarfs. Neutron stars typically have radii $\sim 10$
km and rotation rates up to $\lesssim 1 $ kHz.  Accreting neutron
stars build up very thin oceans (which are, in the zones where burning
and convection occur, composed primarily of hydrogen, helium and
carbon) with $d<10^4$ cm, making the aspect ratio very large
$\eta=r_i/r_o\ge 0.999$.
Convection is expected to be triggered due to thermonuclear burning of
the accreted material, which can take place in an unstable fashion due
to the extreme temperature dependence of the nuclear reactions, giving
rise to the phenomenon of Type I X-ray bursts (see
\citet{Strohmayer06} for a general review of Type I X-ray bursts and
\cite{Malone11,Malone14,Keek17} for more specific discussions of
convection). Surface patterns known as burst oscillations are observed
to develop frequently during thermonuclear bursts, motivating our
interest in convective patterns \cite{Wat12}.
Due to the high densities of matter in the neutron star ocean, the main
contribution to microphysics is from highly degenerate electrons. The
heat capacity can be estimated using the formula for a degenerate gas
of fermions: $ c_P = (\pi^2 k_b / 2 \mu_e m_p ) k_bT / E_F \sim 10^8$
ergs g$^{-1}$ K$^{-1}$\cite{hookhall}. Conductivity, $K \sim 10^{16}$
ergs cm$^{-1}$ s$^{-1}$ K$^{-1}$ , and kinematic viscosity, $\nu \sim
10^6$ cm$^2$s$^{-1}$, are estimated from expressions in
\cite{Yakovlev80,NaPe84}, thus the Prandtl number is of the order $\Pr
\sim 10^2$.
These high viscosities and extremely thin convective layers combine to
ensure that the Taylor number is not so high, ranging from $\Ta=10^8$
for $d=10^4$ cm (the expected ignition depth for Carbon), and $\Ta=1$
for $d=10^2$ cm (the expected ignition depth for hydrogen/helium).
The full accreted ocean extends to a depth of a few times $10^4$ cm.
Given this range of possible values for $\Ta$, very different flow
regimes might be expected, and further research is needed into the
physical conditions of neutron star oceans (our current study does not
fully cover this range of parameter space).  In what follows, and in
Tables \ref{tab2} and \ref{tab3}, we discuss a more restricted range
$\Ta\in[10^4,10^8]$, so as not to extrapolate too far beyond the
bounds our study. 

If we assume $\Pr=10^2$ and $\Ta\in[10^4,10^8]$ to be the valid regime
of a neutron star ocean, it is clear from extrapolating
Fig.~\ref{fig7} that the preferred mode of convection will be SC.  The
$\Pr$ values are beyond the range covered in our study, however the
numerical computations of \cite{Zha92} just fall in to the neutron
star regime, albeit in a thicker shell (geometric effects are believed
to be of secondary importance with respect to $\Pr$). At $\Pr=10^2$
these authors obtained following power laws:
$\Rayc = 1.99 \Ta^{0.66}$, $|\omega_c| = 0.0190 \Ta^{0.33}$ and $m_c =
0.0687 \Ta^{0.16}$ if we use the analytic formulas, and $\Rayc = 4.057
\Ta^{0.66}$, $|\omega_c| = 0.0683 \Ta^{0.33}$ and $m_c = 0.0819
\Ta^{0.16}$ if we use the numerical values at $\Ta=10^8$ given in
Table 2 of~\cite{Zha92} (it should be noted that the definition of
$\Ta$ used in \cite{Zha92} is different to the one used here).  Then,
$\Rayc \sim 10^3-10^6$, $|\omega_c|\sim 10^0-10^1$ and $m_c\sim
10^0-10^1$, which gives rise to a period $P \sim 1-10$ s.

One of the less well understood phenomena occuring on neutron stars
are burst oscillations (for a review see \cite{Wat12}).  These occur
on accreting neutron stars in the aftermath of a Type I X-ray burst,
and are observed as modulation of the X-ray luminosity at a frequency
very close to that of the spin frequency, sometimes drifting by up to
a few Hz in the tail of a burst as the ocean cools. They are caused by
pattern formation in the burning ocean.
Currently the two leading explanations of this phenomena involve flame
spreading
\cite{Spitkovsky02,Cavecchi13,Cavecchi15,Cavecchi16,Mahmoodifar16}, or
global modes of oscillation of the ocean \cite{Heyl04,Piro05}, but no
model yet fits the data precisely, and both mechanisms may be
involved.  Convection is expected to occur in many bursts, however,
and the possible role of convection in pattern formation has yet to be
fully investigated.  With a rotating frame frequency of 0.1 - 1 Hz,
the timescales associated with the convective patterns computed above
are certainly compatible with those required to explain burst
oscillations (where a small rotating frame frequency would required to
keep the observed frequency within a few Hz of the spin frequency).

\begin{figure}[t!]
\begin{center}
\includegraphics[width=0.95\linewidth]{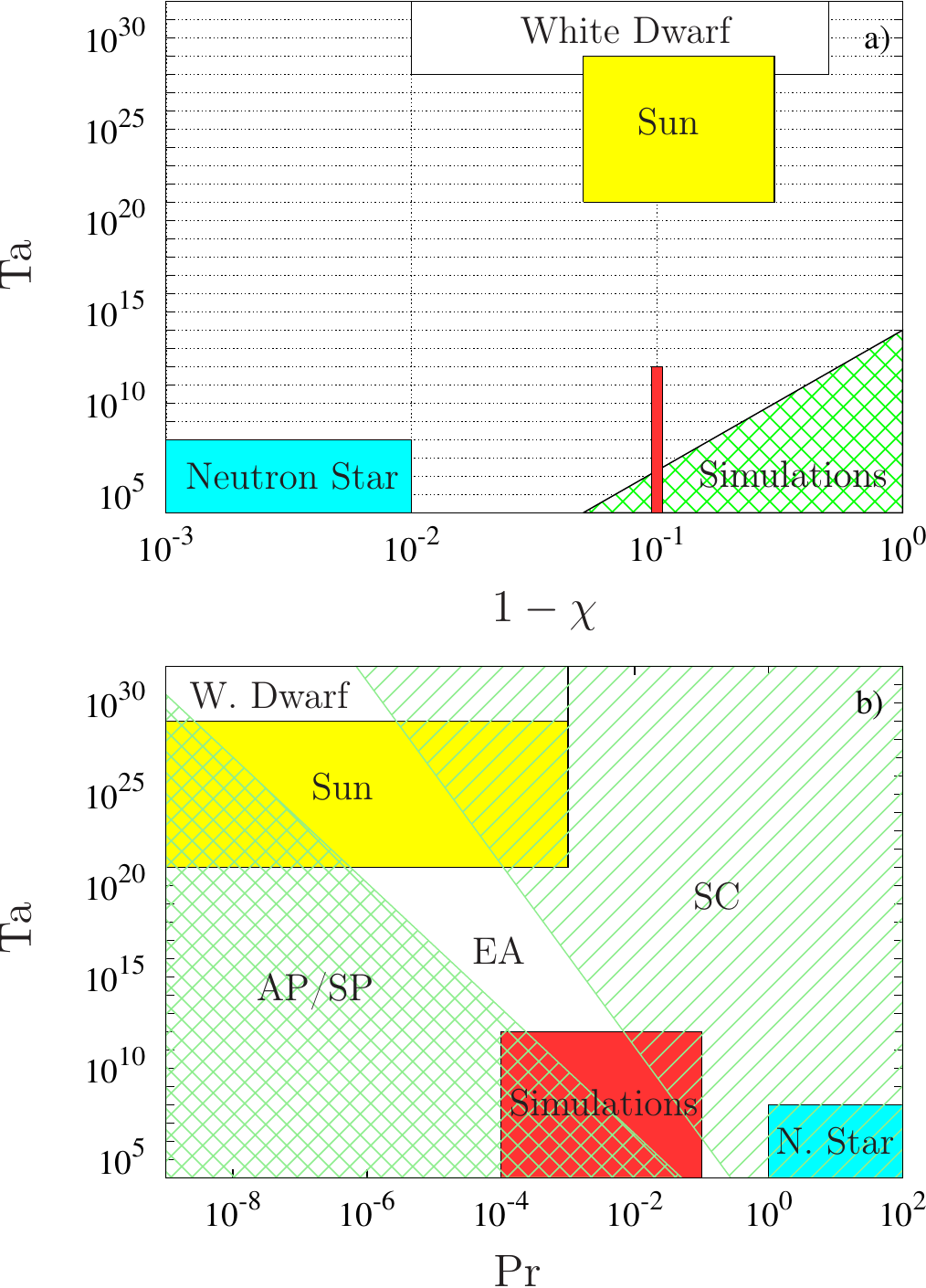}
\end{center}
\caption{a) Estimates of $\Ta$ and $1-\chi$ for stars convective
  oceans and parameter values of the current (red) and previous
  (green) numerical studies of thermal convection which are clearly
  still far from stellar conditions. b) Same as a) but for $\Ta$ and
  $\Pr$. The estimated region of stability of the different convective
  instabilities SC (with stripes), EA (plain) and AP/SP
  (cross-hatched) is also shown.}
\label{fig8} 
\end{figure}

%
%
It should be noted that in these computations we have only accounted for
temperature differences, and do not take into account burning physics
or compositional gradients.  This would not affect estimated
dimensionless parameters $\Pr, \Ta$ but does mean that there is much work
to do before direct comparison with astrophysical phenomena since non-linear effects would become relevant.
To illustrate more easily the parameter regime into which the
astrophysical scenarios mentioned above fall, Fig.~\ref{fig8}(a) and
(b) highlight the region occupied by each (in $(1-\eta,\Ta)$ and
$(\Pr,\Ta)$ parameter space, respectively), together with the region
covered by current and previous numerical studies. In addition,
Fig.~\ref{fig8}(b) includes the region of preference of each
convective instability AP/SP, EA, and SC extrapolated from
Fig.~\ref{fig7}.  While the Sun and white dwarfs have aspect ratios
$\eta$ that are numerically attainable, they have $\Ta$ numbers which
are impossible to handle with current simulations. The latter does not
occur in the case of neutron stars, but in contrast, very thin layers,
demanding prohibitive spatial resolutions, must be considered.

\section{Conclusions}
\label{sec:conc}

The onset of low Prandtl number Boussinesq thermal convection in fast
rotating very thin spherical shells is investigated carefully in
this paper, by means of detailed numerical computations. A massive
exploration of the parameter space $(\Ta,\Pr)$ is performed in a range
of astrophysical interest characterised by low $\Pr$ and high $\Ta$
and very thin spherical shell $\eta=0.9$ with stress-free
conditions. The use of efficient time integration methods has been
useful for integrating the short temporal scales exhibited by the
flows at high $\Ta$ and low $\Pr$. The parallelism of the code allows
us to cope with the high resolutions needed to follow the marginal
stability curves for a wide range of azimuthal wave numbers
$m\in[10,500]$. This is necessary because the curves of quite
different $m$ are very close in the case of very thin shells. In
comparison with previous linear numerical studies in such thin
($\eta>0.8$) geometry~\cite{AHA04}, the present study considers
several orders of magnitude larger $\Ta$ and smaller $\Pr$
numbers. The region of the parameter space covered is even wider than
most of the linear studies in thicker shells. Such very low $\Pr$ and
large $\Ta$ number have previously been reached only in the recent
numerical~\cite{SGN16} and analytic~\cite{ZLK17} studies, and then only for the
case of a full sphere.

A first exploration at four fixed $\Pr<1$ reveals the existence of
three types of preferred modes as $\Ta$ is varied. Prograde spiralling
columnar (SC) convection occurs at larger $\Pr$ and the power laws
obtained for the critical parameters agree quite well with previous
results. For intermediate $\Pr$, equatorial (EA) modes, which can be
retrograde for small $\Ta$, are found to be consistent with former
studies considering thicker shells but, surprisingly, prograde
equatorial antisymmetric or symmetric polar (AP/SP) modes with high
wave number can also dominate at $\Pr=10^{-2}$ and $\Ta=10^{12}$. The
latter type of modes are the only ones preferred at the lowest
$\Pr=10^{-4}$, but in this case they can also be retrograde and show
multicellular structures and strong shear layers which extend in the
latitudinal direction. Antisymmetric polar modes, with convection
confined inside the inner cylinder, were found for first time
in~\cite{GSN08} considering a shell with $\eta=0.4$, but they are
dominant in a substantially smaller range of $\Ta$ when compared with
our results at $\eta=0.9$. The latter fact increases the relevance of
polar antisymmetric convection since low Prandtl number fluids
convecting in very thin shells are common in stellar interiors.

With further exploration, by varying $\Pr$, the transitions among
SC, EA and AP/SP modes are computed and traced in the $(\Ta,\Pr)$
space. This is the first time that this has been done: previous
studies~\cite{ABW97,NGS08} provided only a qualitative sketch in a
smaller parameter region. At the lowest $\Pr$ values, polar modes
become the only ones that are preferred and the critical parameters
become nearly constant, suggesting the zero-Prandtl-limit is not
far. The transition between AP/SP and EA modes takes place at
$\Pr_{\text{P/EA}}=0.68\Ta^{-0.29}$, giving rise to a sharp step in the
drifting frequency $|\omega_c|$ and a jump in $m_c$. The step and
the jump tend to decrease with $\Ta$. Equatorial modes are superseded
by SC at $\Pr_{\text{EA/SC}}=1.67\Ta^{-0.2}$ but only for
$\Ta<2.8\times 10^{9}$.  The transition between EA and SC modes is
also characterised by a sharp step in the drifting frequency
$|\omega_c|$ and a big jump in the azimuthal wave number.

At larger $\Ta$ and moderate $\Pr$ we have found two additional
transitions not described previously.  One is between EA and AP/SP
modes and the other is between AP/SP and SC modes, taking place at
larger $\Pr$. The frequencies $|\omega_c|$ are slightly increased for
the former but strongly decreased for the latter. The situation for
$m_c$ is different: it increases substantially when AP/SP polar modes become
selected but only slightly when SC overcomes the AP/SP modes. The two
transition curves intercept at $(\Ta_3,\Pr_3)=(2.8\times
10^{9},2.12\times 0^{-2})$ giving rise to a triple-point bifurcation that, to the best knowledge of the authors, has never been reported
before in the context of thermal convection in rotating spherical
geometry. At this triple-point, AP/SP, EA and SC modes are dominant and
have very different $|\omega_c|$ and $m_c$. This is relevant because
when nonlinearities are included, a rich variety of chaotic dynamics
may be expected almost at the onset. In addition, as the three types
of modes are characterised by different physical mechanisms, nonlinear
solutions driven by different force balances are expected to coexist
at the triple-point. The study of the associated zonal flows and the
  magnitude of any differential rotation that arise in these different
  nonlinear regimes is important for the understanding of stellar
  ocean convection.

Finally, the fit formulae computed are used to estimate the critical
parameters, the characteristic time scales and the most likely mode
for the onset of convection occurring in the Sun, white dwarfs and
accreting neutron stars. Although Boussinesq thermal convection in thin
rotating spherical shells fails to reproduce the compressibility
effects in stellars fluids, it captures the essential features of rotation
and spherical geometry and thus gives valuable insight for further
studies of more realistic models. Using known values of the physical
properties, reasonable results for the critical Rayleigh number and
the time scales are achieved, which are of similar order to
observational phenomena reported in the literature.

According to our results equatorial modes (and polar modes with less
degree) are the best candidates in the case of Sun and cooling white dwarf
convection because of their low
$\Pr$. Their time scales are about $10^2-10^4$ years for the Sun,
encompassing the well-known long term periods Gleissber and Suess or
Hallstatt cycles, and slightly lower at ($10^2-10^3$) years for
white dwarfs.  In the case of accreting neutron star oceans
the situation seems to be quite different. It is not clear which
regime they will belong to, because they have very viscous (electron
degeneracy) fluids. Our results suggest $\Ta=10^0-10^8$ depending on
the size of the shell considered and $\Pr=10^2$. If we assume
$\Ta>10^3$ then convection will take the form of spiralling columns
drifting very rapidly (on time scales of $1-10$ seconds). These timescales
are consistent with some of the timescales seen in thermonuclear bursts where the accreted ocean develops patterns known as burst oscillations, and may therefore be of interest to studies of this as yet unexplained phenomenon.

\section*{Acknowledgements}
The authors acknowledge support from ERC Starting Grant No. 639217
CSINEUTRONSTAR (PI Watts).  They also wish to thank Marta Net, 
 Juan S\'anchez, Sumner Starrfield and David Arnett for their helpful suggestions and comments.

\clearpage
\newpage
\onecolumngrid
\section*{Erratum: The onset of low Prandtl number thermal convection in thin
  spherical shells [Phys. rev. Fluids 3, 024801 (2018)]}
\twocolumngrid

\begin{table}[t!]
  \begin{center}
  \begin{tabular}{llcccc}
  \hline\\
&  Property               & Accreting neutron   \\
&                        &  star ocean \\[3pt]
  \hline \\
& $\kappa$ (cm$^2$s$^{-1}$) &  $10^{3}-10^{5}$ \\  
& $\nu$ (cm$^2$s$^{-1}$)    &  $10^0$          \\
& $d$ (cm)                 &  $10^{2}-10^{4}$  \\
& $\Omega$ (s$^{-1}$)       &  $10^{2}$  \\
  \\
  \hline
\end{tabular}
\end{center}                                                                                                                         
\caption{Physical parameters corresponding to an accreting neutron
  star ocean. Thermal diffusivity $\kappa$, kinematic viscosity $\nu$,
  layer width $d$ and rotation rate $\Omega$.  Values are taken
  from~\cite{Yakovlev80bb,NaPe84bb,Wat12bb}.}
\vspace{-4.mm}
\label{tab1err}
\end{table}

We have performed the linear stability analysis in thin rotating
spherical shell thermal convection models at low Prandtl number. The
results have been used to extrapolate to several astrophysical
scenarios, namely, the sun, cooling white dwarfs and accreting neutron
stars.  Unfortunately there was an error in the estimation of the
kinematic viscosity for the case of accreting neutron star oceans,
which we correct here. This erratum concerns only a small part of the
paper in Sec. VI involving one column in Tables II and III, and
Fig. 8. The changes are discussed in the following.

Tables~\ref{tab1err} and~\ref{tab2err} (corresponding to Tables II and III
of the original paper) provide the correct values for the accreting
neutron star ocean scenario. A valid estimation for the kinematic
viscosity $\nu$ is $\nu \sim 1$ cm$^2$ s$^{-1}$ rather than $\nu \sim
10^6$ cm$^2$ s$^{-1}$ stated in the original paper. The new value of
$\nu$, and slightly amended values for the thermal conductivity
$\kappa=10^3-10^5$ cm$^2$ s$^{-1}$ and gap width $d=10^2-10^4$ cm
($\kappa=10^6$ cm$^2$ s$^{-1}$ and $d=10^3-10^4$ cm were supposed in
our paper) give rise to new estimations of the parameters [Prandtl
  number ($\Pr$) and Taylor numbers ($\Ta$)], critical parameters
[critical Rayleigh number ($\Rayc$), wave number $m_c$, and frequency
  $\omega_c$] and typical time scale $P=(2\pi
m_c/|\omega_c|)d^2/\nu$. They are shown in Table~\ref{tab2err}.

\begin{table}[b!]
  \begin{center}
  \begin{tabular}{lllcccc}
  \hline
& Type of mode &  Property        & Accreting neutron  \\
&              &                  &  star ocean \\[3pt]
\hline                                                                                              \\
&              & $\Ta^{}$          &   $10^{12}-10^{20}$       \\
&              & $\Pr^{}$          &   $10^{-5}-10^{-3}$       \\  
&              & $\Pr_{\text{P/EA}}$  &   $10^{-4}-10^{-6}$       \\
&              & $\Pr_{\text{EA/SC}}$ &  $10^{-2}-10^{-4}$        \\
\\
\hline
\\
&              & $\Rayc$         &   $10^{6}-10^{10}$       \\
&              & $|\omega_c|$    &   $10^{4}-10^{7}$       \\
& SP           & $m_c$           &   $10^{2}-10^{4}$       \\
&              & $P$ (years)     &   $10^{-5}-10^{-2}$       \\
\\
\hline
\\
&              & $\Rayc$         &  $10^{5}-10^{10}$      \\
&              & $|\omega_c|$    &  $10^{5}-10^{9}$      \\
& EA           & $m_c$           &  $10^{2}-10^{3}$      \\
&              & $P$ (years)     &  $10^{-6}-10^{-5}$       \\
\\
\hline
\\
&              & $\Rayc$          &  $$       \\
&              & $|\omega_c|$     &  $10^{6}-10^{9}$        \\
& P            & $m_c$            &  $10^{1}-10^{3}$        \\
&              & $P$ (years)      &  $10^{-7}-10^{-6}$        \\
\\
  \hline
\end{tabular}
\end{center}                                                                                                                         
\caption{Estimation of the critical parameters and a typical time
  scale $P=(2\pi m_c/|\omega_c|)d^2/\nu$ in seconds (this is the
  timescale associated with the drifting phase velocity as computed in
  the rotating frame) for the Sun, a white dwarf and a neutron star
  depending on the type of mode, spiralling (SC), equatorial (EA) or
  polar (P). All of them could be preferred depending on appropriate
  particular choice of $\Pr$. This is indicated by the critical $\Pr$
  at the P/EA and EA/SC transitions, which are also shown.}
\label{tab2err}
\end{table}

\begin{figure*}[t!]
  \begin{center}
\vspace{3.mm}
\includegraphics[width=0.4\linewidth]{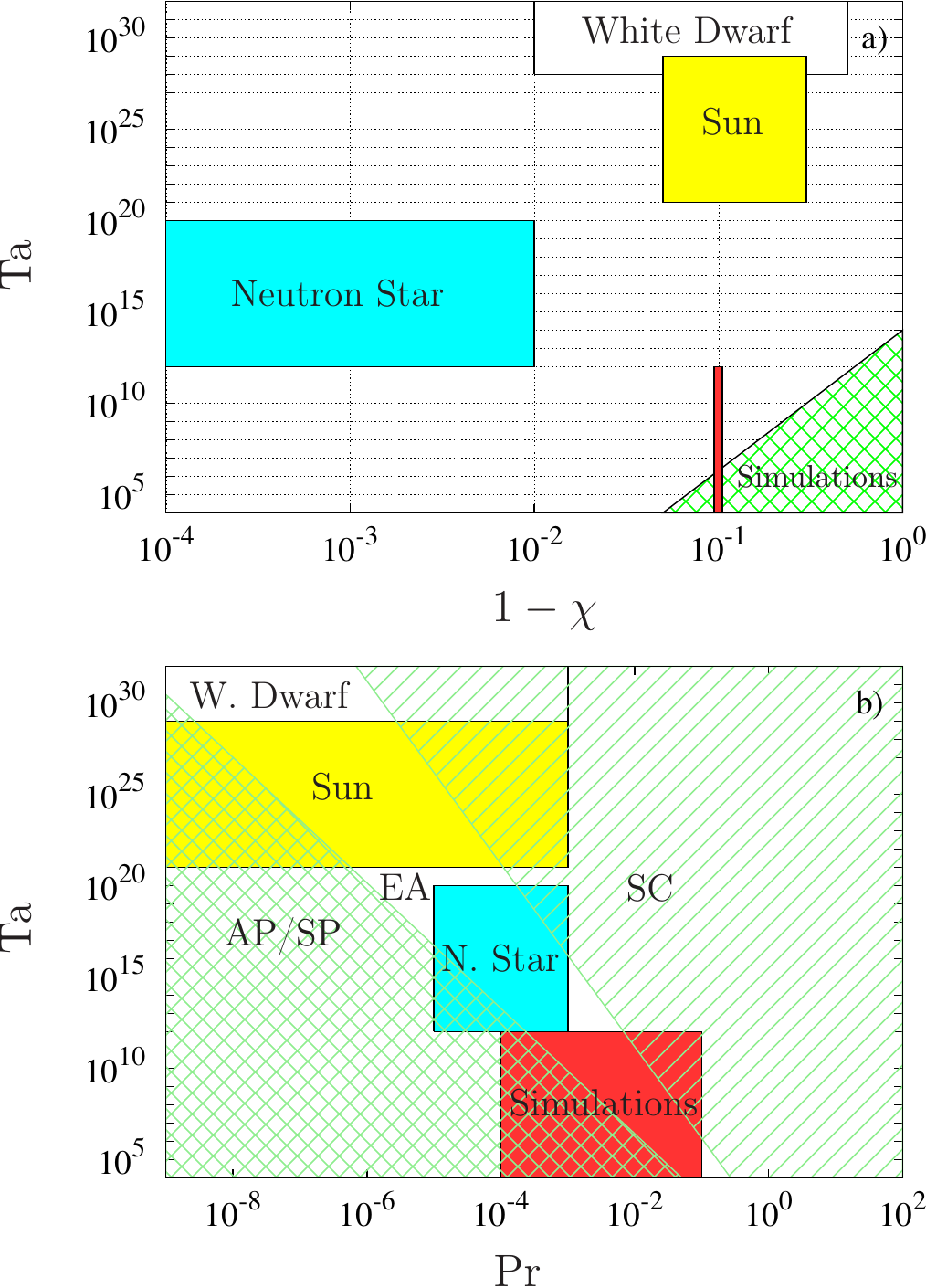}
\end{center}
\caption{a) Estimates of $\Ta$ and $1-\chi$ for stellar convective
  oceans and parameter values of the current (red) and previous
  (green) numerical studies of thermal convection, which are clearly
  still far from stellar conditions. b) Same as a) but for $\Ta$ and
  $\Pr$. The estimated region of stability of the different convective
  instabilities SC (with stripes), EA (plain) and AP/SP
  (cross-hatched) is also shown.}
\label{fig1err} 
\end{figure*}

As a result of the new viscosity the Prandtl number of an accreting
neutron star ocean becomes very low $\Pr\in[10^{-5},10^{-3}]$ and the
Taylor number moderately large $\Ta\in[10^{12},10^{20}]$, and thus our
linear stability analysis falls quite close to this regime. This is
shown in Fig.~\ref{fig1err} (corresponding to Fig. 8 of our paper). At
this regime, according to our extrapolation in the original paper for
the critical $\Pr$ marking the transition among modes the most
feasible mode is of equatorial (EA) type instead of spiralling (the
result reported in the original paper).  However it may also be of
polar (P) type, due to the new estimation of very low $\Pr$ and
moderately large $\Ta$.

Comparing the new Table~\ref{tab2err} (EA and P modes) with Table III of
our paper, the new critical parameters and time scales have not
increased substantially. For instance, we have obtained time scales on
the order of 10-100 s for EA modes, and on the order of 1-10 s for P
modes. These are similar to those obtained in the original paper for
spiralling modes at $\Pr=10^2$ using the estimations provided
in~\cite{Zha92bb}, but this should not be surprising because two
parameters ($\Pr$ and $\Ta$) have changed.

The main conclusion of this erratum is that accreting neutron star
oceans are characterized by low Prandtl numbers, rather than larger
ones, as for the cooling white dwarfs and the sun scenarios. Thus low
Prandtl numbers in thin shells (the regime our paper addresses) should
be considered for their modeling. Then, because the main difference
between neutron star oceans and white dwarfs or solar convective zones
is now their width, different $\Ta$ regimes, giving rise to very
different time scales, should be expected. As concluded
in our original paper these time scales could be linked with some
observational properties in the three astrophysical scenarios
considered.

%


\end{document}